\documentclass[conference]{IEEEtran}
\IEEEoverridecommandlockouts
\usepackage{cite}
\usepackage{amsmath,amssymb,amsfonts}
\usepackage{algorithmic}
\usepackage{graphicx}
\usepackage{textcomp}
\usepackage{xcolor}
\usepackage{caption}
\usepackage{subfigure}
\def\BibTeX{{\rm B\kern-.05em{\sc i\kern-.025em b}\kern-.08em
    T\kern-.1667em\lower.7ex\hbox{E}\kern-.125emX}}
\begin{document}

\title{Network2Vec: Learning Node Representation Based on Space Mapping in Networks}



\author{\IEEEauthorblockN{Zhenhua Huang}
\IEEEauthorblockA{\textit{School of Software Engineering} \\
\textit{South China University of Technology}\\
zhhuangscut@gmail.com}

\and
\IEEEauthorblockN{Zhenyu Wang}
\IEEEauthorblockA{\textit{School of Software Engineering} \\
\textit{South China University of Technology}\\
wangzy@scut.edu.cn}

\and
\IEEEauthorblockN{ Rui Zhang}
\IEEEauthorblockA{\textit{School of Software Engineering} \\
\textit{South China University of Technology}\\
zhang1rui4@outlook.com}

\and
\IEEEauthorblockN{Yangyang Zhao}
\IEEEauthorblockA{\textit{School of Software Engineering} \\
\textit{South China University of Technology}\\
msyyz@mail.scut.edu.cn}

\and
\IEEEauthorblockN{ Xiaohui Xie}
\centering
\IEEEauthorblockA{\textit{Dept. of Computer Science} \\
\textit{University of California, Irvine}\\
xhx@uci.edu}

\and
\IEEEauthorblockN{ Sharad Mehrotra}
\IEEEauthorblockA{\textit{Dept. of Computer Science} \\
\textit{University of California, Irvine}\\
sharad@uci.edu}
}

\maketitle

\begin{abstract}
Complex networks represented as node adjacency matrices constrains the application of machine learning and parallel algorithms. To address this limitation, network embedding (i.e., graph  representation) has been intensively studied to learn a fixed-length vector for each node in an embedding space, where the node properties in the original graph are preserved. Existing methods mainly focus on learning embedding vectors to preserve nodes proximity, i.e., nodes next to each other in the graph space should also be closed in the embedding space, but do not enforce algebraic statistical properties to be shared between the embedding space and graph space. In this work, we propose a lightweight model, entitled Network2Vec, to learn network embedding on the base of semantic distance mapping between the graph space and embedding space. The model builds a bridge between the two spaces leveraging the property of group homomorphism. Experiments on different learning tasks, including node classification, link prediction, and community visualization, demonstrate the effectiveness and efficiency of the new embedding method, which improves the state-of-the-art model by 19\% in node classification and 7\% in link prediction tasks at most. In addition, our method is significantly faster, consuming only a fraction of the time used by some famous methods.
\end{abstract}


\section{Introduction}
Many systems are composed of complex networks or graphs, such as social networks, biological networks, co-authorship network, web page networks and communication networks \cite{evolution2007graph, wikicollective2008data, blogflickr2009}. Nodes or vertices in these networks are entities and edges that denote relations or interactions between nodes. Traditional methods to represent the networks are based on a sparse adjacent matrix, in which nodes are represented as vectors in which only the adjacent relations between nodes are kept without semantic meanings. Therefore, many scholars dedicated on representing nodes by low-dimensional dense vectors, in which some hidden rich semantic attributes about the nodes are involved, e.g., the social influence, roles or difference of the nodes in a network \cite{survey2017comprehensive, survey2017network}. While specific and well-designed algorithms are required to calculate those attributes before network embedding methods, e.g., using PageRank to calculate social influence.

\begin{figure}
\centering
\includegraphics[width=3.0in, height=1.5in]{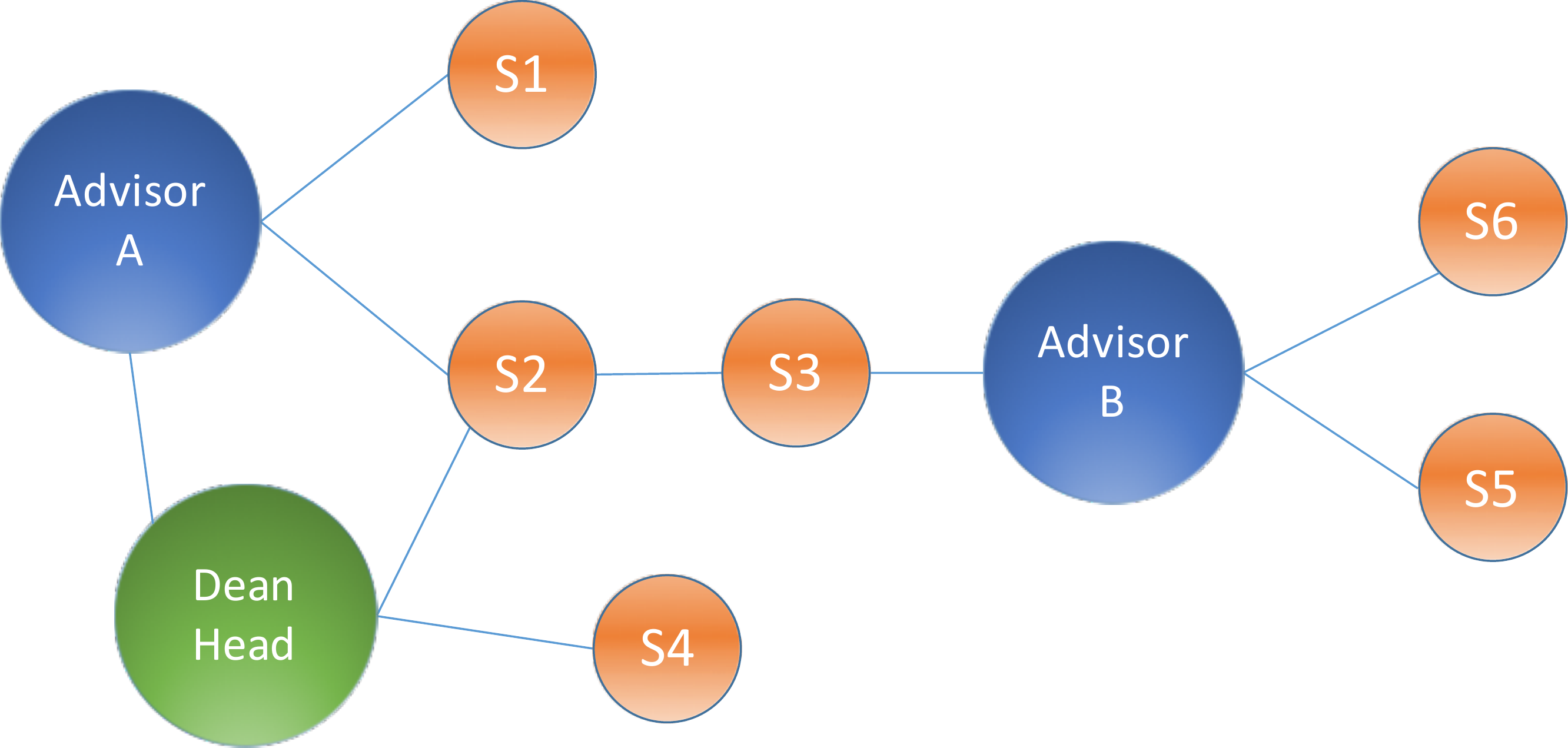}
\caption{Attributes beyond neighbors.} \label{cross_neigh}
\end{figure}

\textbf{Network Embedding}, also referred to as \textbf{Graph Representation} or \textbf{Network Representation}, can be interpreted as a transformation or fully mapping between the physical graph space and embedding vector space. After the mapping, the characteristics of nodes and edges in the original physical space are expected to be preserved as much as possible in the embedding vector space. Many models have been proposed to embed nodes in networks, such as DeepWalk\cite{Deepwalk2014}, LINE\cite{line2015}, GrapRep\cite{Grarep2015}, Node2vec\cite{node2vec2016} etc. Although many network embedding methods based on stochastic gradient descents have been proposed, the relationship between embedding vectors space and physical space or topological properties are not highly discovered and remain opaque. One direct way of graph representation is building a 'bridge' function that connects the two spaces and constructing some global statistical indicators in graph space which can reflect similarity and distinction between nodes.
Motivated by the idea, we discover the relationship between the two spaces and propose a novel light-weight and flexible method, \textbf{\emph{Network2Vec}}, to learn embedding vectors by directly space mapping. The model applies a homomorphism mapping to learn the relationship of elements in the embedding space and statistical indicators in the physical space that contains global and local information. In some cases, the nodes with similar attributes are even not directly connected. For example, in Fig. \ref{cross_neigh}, the advisor $A$ and advisor $B$ are professors, and the student $S_2$ and node $S_3$ are students. Although the advisor $A$ and $B$ are not one-hop neighbors, they share the same attribute of the teacher. Therefore, we design a jumping mechanism to address the problem and improve the flexibility of the model. The model is conceptually simple but empirically powerful. Experimental results in different tasks and datasets have shown that the model is competitive and effective compared with previous models.

The contributions of the paper are in the following:
\begin{enumerate}

\item We have studied the relationship between the graph space and embedding space in network embedding using group homomorphism theory. It is found that there is a log-linear mapping relationship between the elements in the two spaces. Therefore, a new light-weight model $Network2vec$ is proposed to train the embedding vectors in an efficient way. The model has a low time complexity and runs faster than most existing models while maintains a competitive performance in different tasks.

\item We point out the pure matrix decomposition on occurrence matrix can not achieve better results than DeepWalk \cite{Deepwalk2014} in network embedding due to noisy impacts, although they are theoretically equal. A log penalty function is applied to restrain the noisy data and power-law influences. 

\item We optimize the way of random walks and calculating statistical indicators so that the model is able to perceive similarities and differences between nodes in graph space and take both local and global information into consideration, which facilitates producing a better node representation. The model performs well in node classification and link prediction, improving the baseline models by a relatively large margin.
\end{enumerate}

\section{Related Works}
The first generation methods of network representation include graph Laplacian Eigenmaps \cite{laplacian2002}, and Spectral Clustering \cite{spetralclustering2011}, which mainly focus on dimension reduction and network reconstruction without network inference thus rich structures and properties are not incorporated. DeepWalk \cite{Deepwalk2014} is a remarkable method that generalizes the language representation model skip-gram \cite{word2vec2013efficient} into node representation. The method has a better performance than Spectral clustering method on the multi-class classification task. The LINE \cite{line2015} carefully defines the concept of first-order and second-order similarity, which has a high motivation for our paper. Both the works have drawn much attention over last several years. One extension work of LINE is SDNE \cite{SDNE2016structural}, which applies a deep auto-encode network for the first time in learning the structural similarities. While the Grover et al. \cite{node2vec2016} believe the diverse random walks benefit the scalability embedding model and generalizes the DeeWalk to node2vec. NetMF\cite{netmf2018wsdm} unifies DeepWalk, Node2vec, and LINE as matrix factorization problems and improve the embedding by graph theory. Feng et al. \cite{aaai2018scale} found the power-law distribution has significant impacts on node embedding learning and proposed a method by constraining vertex embedding and improving random walk method. The above methods are classified as \textbf{Unsupervised Network Embedding}. Nodes in the real-world networks are sometimes accompanied by abundant external information such as textual information, tags, etc. It motivates some works to integrate external information to enhance the effectiveness of representation vectors, such as TADW \cite{TADW2015network}, and GCN \cite{semiGCN2016semi}. While some works mainly concern network presentation targeting at certain tasks such as classification \cite{MMDW2016max}, which can be categorized as \textbf{\emph{Task-Oriented Network Embedding}}. Besides, many scholars focus on using graph neural networks to improve network representation. While our paper focuses on unsupervised network embedding.

Even though some works \cite{TADW2015network} believe the DeepWalk equals to matrix decomposition in theory, while in fact, the pure matrix factorization cannot produce results as well as DeepWalk. Even in NetMF, value constrains method is applied to produce a better performance. Different from NetMF \cite{netmf2018wsdm} that is based on spectral graph theory and learn all the nodes embedding together, we point out the importance of statistical regularities between embedding space and graph space, and build a bridge between two spaces, making the model more scalable and reasonable in mathematical theory. And the matrix can be stored in a sparse matrix which saves memory space. Different with DeepWalk \cite{Deepwalk2014}, our statistical indicator is more flexible with both local and global network information. Furthermore, our works focus on revealing the relationship between the embedding space and graph space.



\section{Network2vec}
\begin{figure}
\centering
\includegraphics[width=2.5in, height=2.5in]{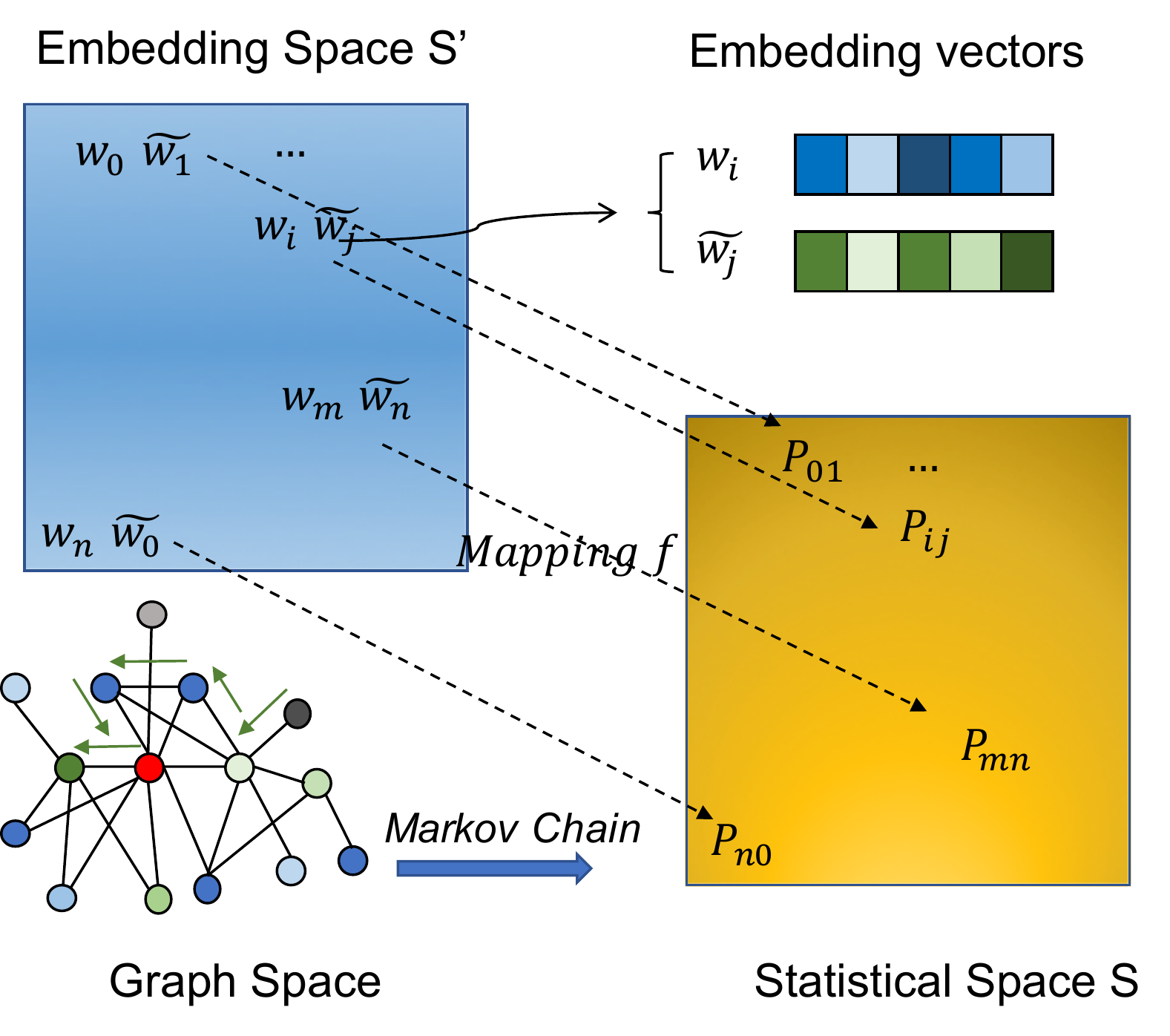}
\caption{The space mapping of the Network2vec.} \label{fig1}
\end{figure}
\subsection{Preliminaries}\label{AA}
Given a network $G$ = $(V,E)$, where $V$ is the vertex set and $E$ is the edge set. The network embedding task is to learn an embedding vector for each node so that the properties of nodes in graph space are kept in the vector. Let $f$ be the mapping function that describes the relationship between nodes in the embedding space and the graph space in networks. Here, we denote $w_i$ as a vector when node $i$ is the center node and $\widetilde{w_i}$ as context node, similar to \cite{word2vec2013efficient}.
Suppose the similarity function in embedding space is $S'(i,j)$. We use $S'(i,j)=w_i^T w_j$, ($|w_i|$=1.0), to represent the distance or similarity of node $i$ and node $j$ in embedding space. And $w_i^T(\widetilde{w_j} - \widetilde{w_k})$ donates the difference between node $j$ and $k$ to node $i$.
We can apply some strategies (e.g., random walks or spectral theory) to calculate statistical indicators that reflect the similarity and difference between nodes in graph space. The calculated indicator matrix is donated as space $S$, which may have many kinds of forms. The statistical space $S$ builds a bridge between the physical graph space and the Euclidean embedding space.

Then, in the following work, all we have to do is find a function $f$ that satisfies the following expectation:
$ f(S'(i,j)) = S(i,j)$. In the function, each node pair in $S‘$ will be mapped by the function $f$ and corresponds to a unique value in another space $S$. According to previous works, several candidate indicators that can be used to measure the node similarity in the physical space are Point-wise Mutual Information \cite{pmi2014nips}, and Co-occurrence probability \cite{glove2014}, and normalized Laplacian matrix \cite{netmf2018wsdm}.

\subsection{Proofs and Transformation}

The distance of two elements in the embedding space $S'$ is represented as  $f(S'(i,j)-S'(i,k)) = f(w_i^T(\widetilde{w_j} - \widetilde{w_k}))$.  

However, the difference between two elements in space $S$ is not measured by minus operation in one space, but often in the operation of division since the indicators are probabilities according to previous studies \cite{glove2014, pmi2014nips}. The following equation is expected to exist between the two spaces:
$ f(S'(i,j)-S'(i,k)) = S(i,j) - S(i,k) = S(i,j) / S(i,k) $

In the space $S'$, the elements are in the form of $w_i w_j$, the operation of difference is minus '$-$', while in the space $S$, the elements are $P_{ij}$, and the operation of difference is dividing '$/$'. Thus, the mapping has a property of $f(a-b)$ = $f(a)/f(b)$. It reminds us the there might be a group homomorphism between the two spaces. Suppose there is a function $f$, making the group homomorphism exists between the two spaces. Then we try to solve the function $f$ and verify the hypothesis by group theory.

There might be several forms of $f$, but one simple solution is $f = exp$.  When $f = exp$, we can verify that the physical space and the embedding space is a group homomorphism between groups  (R; -)  and  (R$>$0; /). 

\textbf{Proofs and Analysis:}\\
\indent $\forall i, j \in N:$  we\quad have, \quad $f(S'(i,j)) = S_{ij} = P_{ij}$,  fully\quad mapping between the space $S'$ and $S$.\\
\indent $\forall i, j, k \in N:$
\begin{equation}
\begin{aligned}
\indent f(S'(i,j)-S'(i,k)) &= exp(w_i^T\widetilde{w_j} - w_i^T\widetilde{w_k})\\
 &= exp(w_i^T\widetilde{w_j})/ exp(w_i^T\widetilde{w_k})\\
  &= P_{ij}/P_{ik} = f(S(i,j))/f(S(i,k))
\end{aligned}
\end{equation}
\selectfont
\par
From the above proof, we can say, for each element in the space $S'$, we can find a corresponding element in the space $S$. And for each element pair $<i,j>$ in the space $S'$ and operation $*$, there is a unique corresponding element pairs operation $*'$. The space $S$ and $S'$ is a group homomorphism by the function $f$. Therefore the above assumption is proved.

Then we use some indicators to specify the model. If the indicator is co-occurrence probability, the final objective function is:
\begin{equation}
  L=\sum_{m\in V, k \in N(m)}||w_m^T\widetilde{w_k}-\log(P_{mk})||
\end{equation}
Here $N(m)$ is the neighbors of node $m$ , in which $P_{mk}$ is probability node $k$ occurs in the context of node $m$. The form seems like a matrix factorization on $S$. We can directly optimize it and mark the method as baseline model $MF$. However, directly optimizing the equation can not achieve a high performance as well as DeepWalk\cite{Deepwalk2014} due to much noisy data in calculate indicators in the space $S$. So $p_{mk}=log(1+C_{mk})$, a penalty function is applied to address the problem, wherein the $C_{mk}$ is the frequency of node $m$ and node $k$ occurs in the same walks. 
\begin{equation}
  L=\sum_{m\in V, k \in N(m)}p_{mk}||w_m^T\widetilde{w_k}-\log(P_ {mk})||
\end{equation}
We use $b_m$ to learn the $log\,C_m$ in the function and one more bias $\widetilde{b_k}$ for context node. The bias is optional according to indicators $S_{mk}$. If bias is concluded, the derivative is:
\begin{equation} 
    \frac{\partial L}{\partial w_m}=\sum_{k \in N(m)}p_{mk}\widetilde{w_k}(w_m^T\widetilde{w_k}-\log(P_{mk})) 
\end{equation}

If the indicator is PMI \cite{pmi2014nips}, $ S_{mk}= log\frac{\#(m,k)|D|}{\#(m)\#(k)}-log\,n$, where $D$ is the corpus and $n$ is the negative samples,  $\#(m,k)$ is the joint probability of node pairs, and $\#(m)$, $\#(k)$ are the frequencies that they occur independently. Since the previous work \cite{pmi2014nips} has demonstrated the model is theoretically equal with Skip-gram, which is the same with DeepWalk \cite{Deepwalk2014} and Node2vec \cite{node2vec2016}, the experiments are ignored here. Our method can also be roughly regarded as an improvement over matrix factorization to some extent, but we have derived the process of network representation and studied relationships between graph space and embedding space, and we decompose a different matrix with the method as described in \cite{TADW2015network}. 
\subsection{\textbf{Statistical Indicators}}

\begin{figure}
\centering
\includegraphics[width=3.0in, height=1.25in]{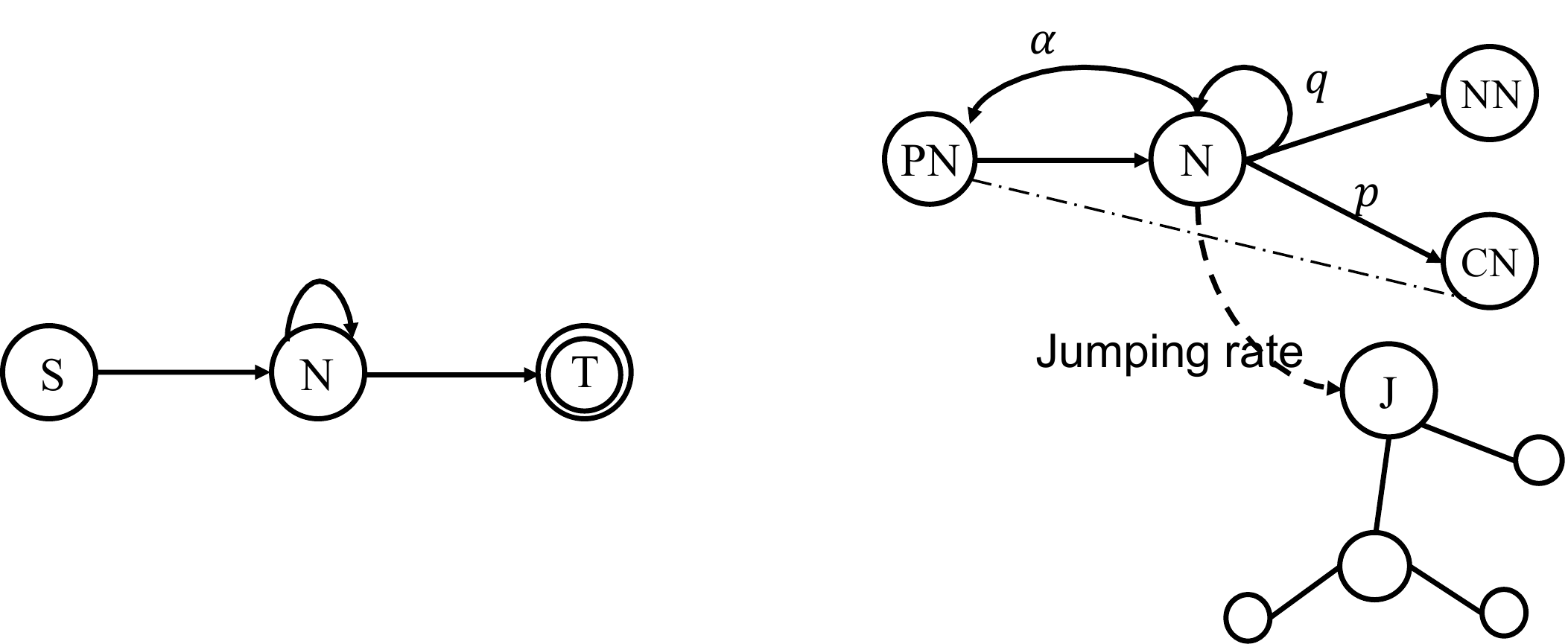}
\caption{Left:the first-order Markov random walk. Right: the second-order Markov random walk with jumping mechanism.}
\label{markov}
\end{figure}

To train our model, the statistical indicators (e.g., co-occurrence matrix and Point-wise Mutual Information (PMI) matrix) need to be calculated. One simple method to estimate the indicator is to generate random walks for nodes based on a first-order Markov Chain and then calculate the frequency and joint probabilities of nodes occur in the same path \cite{Deepwalk2014}. Here, the frequency is weighted by a distance $1.0/dist$ in a path so that the local structure information can be taken into consideration, wherein $dist$ is the distance between the central node and the context node. In the first-order Markov Chain, the current state will choose a neighbor randomly without considering the previous state shown in Fig. \ref{markov}, wherein the $S$ represents start state and the $T$ represents end state. A more generalized method is walking under the search probability as a second-order Markov Chain motivated by \cite{node2vec2016}. As shown in Fig.\ref{markov}, when deciding which neighbors to be selected, the current state will calculate a probability matrix to give a weight for different linked edges. In Fig.\ref{markov}, $CN$ represents common neighbors of both current and previous node with a weight $q$, $NN$ means new neighbors of current and previous node with a weight $\alpha$. The current state can also go to previous state with a weight of $p$. The state transition probabilities are calculated based on : $weight/ (p*|NN|+q*|CN|+\alpha)$. To allow the node to perceive more global network information, the node can jump into other nodes with similar structure in Markov random walks. We divide the degree into 100 segments and map the nodes into their corresponding segments. Then feature of a node and its neighbors' degree distribution are calculated. Then the Jensen–Shannon divergence \cite{manning1999jsd} are used to calculate similarities between nodes. We randomly select one of the most similar nodes by roulette to do jumping. The jumping rate is chosen between {0.0, 0.03, 0.05, 0.1}. In the jumping mechanism, nodes with similar structure role are more likely to be sampled together. We mark the first-order Markov Chain as $N2V_{1nd}$, in which the current state of random walking is not related with previous states. By the jumping mechanism, the nodes sharing similar structure have a higher probability to be sampled together. The mechanism facilitates the model by obtaining more global semantic information of the network. The second-order Markov Chain is marked as $N2V_{2nd}$, in which the next state is related with both current and previous states. 


\subsection{\textbf{Complexity}}
The complexity of the training process depends on the non-zero values of node co-occurrence or PMI matrix. For real-world large networks, the networks are usually sparse and the node degree follows a power-law distribution. Thus, the computation complexity is $K*|V|$ in most cases, $K$ is a constant (i.e., the average number of co-occurrence nodes), better than the worst case $|V|^2$. Compared with DeepWalk \cite{Deepwalk2014}, Node2vec \cite{node2vec2016}, LINE \cite{line2015} and NetMF \cite{netmf2018wsdm}, our model is scalable and time saving.

\section{Experiments}
In this session, we will verify the effectiveness and efficiency of the $Network2Vec$ on the tasks of node classification, link prediction and community visualization. The number of walks and walk length for a node are set to 10 and 80, respectively. The optimizing method is AdaGrad with a learning rate of 0.01 and 0.03. For $N2V_{2nd}$, the $p$ and the $\alpha$ is fixed to 1.0 and $q$ is chosen from $\{1.0, 0.5, 0.25 \}$. In all the models, the embedding size is set to 128. In DeepWalk \cite{Deepwalk2014} and Node2vec \cite{node2vec2016}, the walk length and number of walks are the same with $N2V$. And for Node2vec \cite{node2vec2016}, $p$ and $q$ are also chosen among $\{1.0, 2.0, 4.0\}$ for fair comparison. For LINE \cite{line2015}, we use the default parameters in the paper. The NetMF \cite{netmf2018wsdm} has a high memory cost and have a memory error in the large network like the Flickr Network, so the results of NetMF are not shown in the comparison experiments.

\subsection{\textbf{Node classification}}
To evaluate the effectiveness of network embedding methods in node classification, five data sets are listed as in Table \ref{graphdatasets}. The \textbf{Email} \cite{evolution2007graph} , \textbf{BlogCatalog} \cite{blogflickr2009}, and \textbf{Flickr} \cite{blogflickr2009} networks are commonly used graph benchmarks with ground truth. A multi-class classifier \cite{liblinear2008} is trained based on one-vs-rest logistic regression with different training ratios on the three datasets. The \textbf{Cora} network consists of 5429 links and 2708 scientific publications that are classified into one of seven areas. Motivated by the previous works \cite{yang2015rain}, we label the data according to whether the nodes are the leaders in their research areas, or the structural holes between different domains. The \textbf{Airport} \cite{struc2vec} network is an undirected network where nodes correspond to airports and edges indicate the existence of commercial flights in Brazil. Airports are assigned a label corresponding to their levels of activity. 
\begin{table}
\centering
\caption{Graphs used in node classification.}

\begin{tabular}{l| c | c|c |c | c } \hline
& Email &	 BlogCatalog & Flickr & Airport &Cora   \\ \hline
$|$V$|$& 2405& 10,312 & 80,513 & 131 & 2708  \\ \hline
$|$E$|$& 17981&	333,983 & 5,899,882 & 1074 & 5278  \\ \hline
$|$y$|$& 42&	39&	     195 & 4 & 3  \\ \hline
Label&  Category	&Interests	&Group & Activity& Structure \\ \hline
\end{tabular}
\label{graphdatasets} 
\end{table}


\begin{figure}[]
\centering
  \subfigure{
    \includegraphics[height=1.2in, width=1.6in]{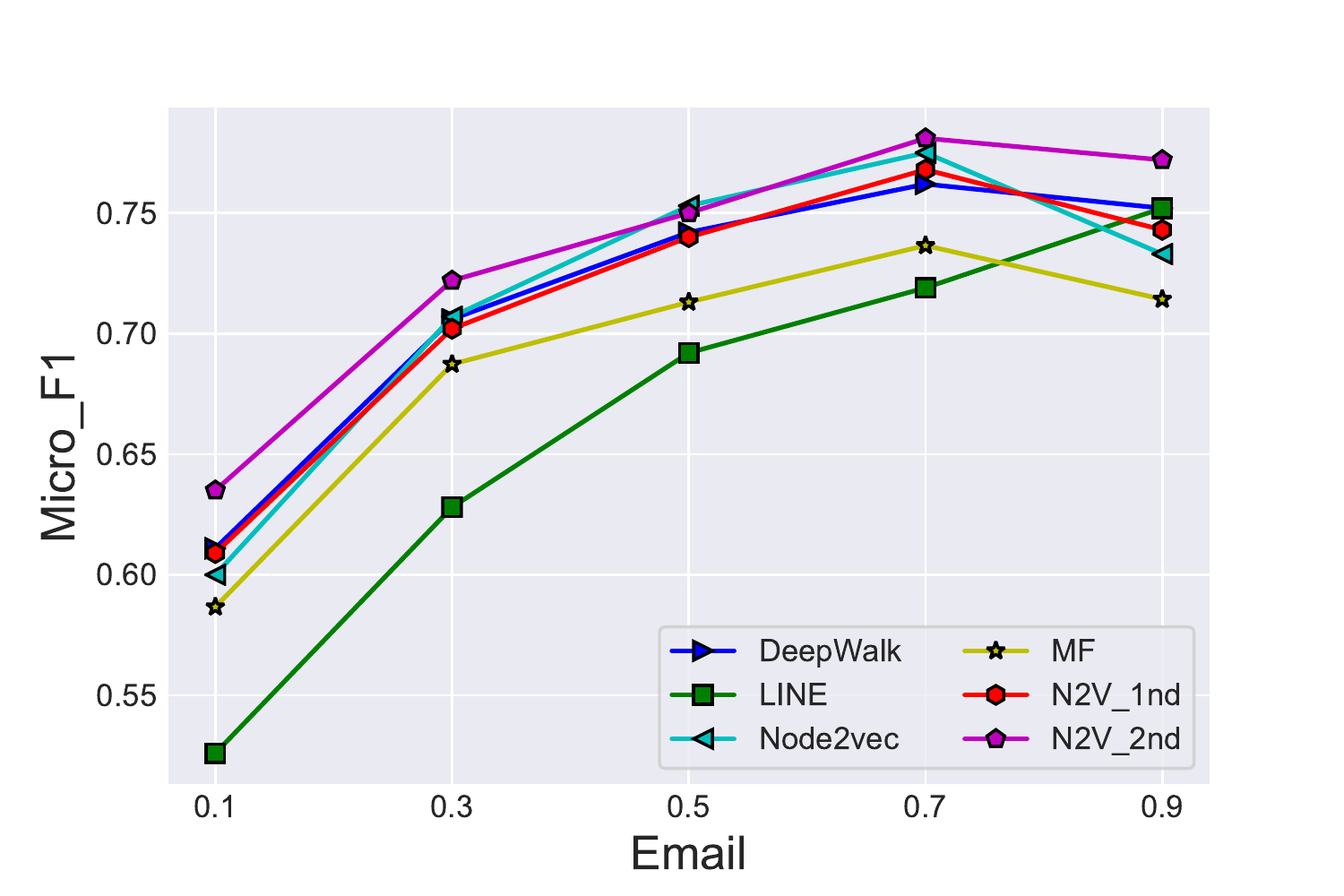}
    \includegraphics[height=1.2in, width=1.6in]{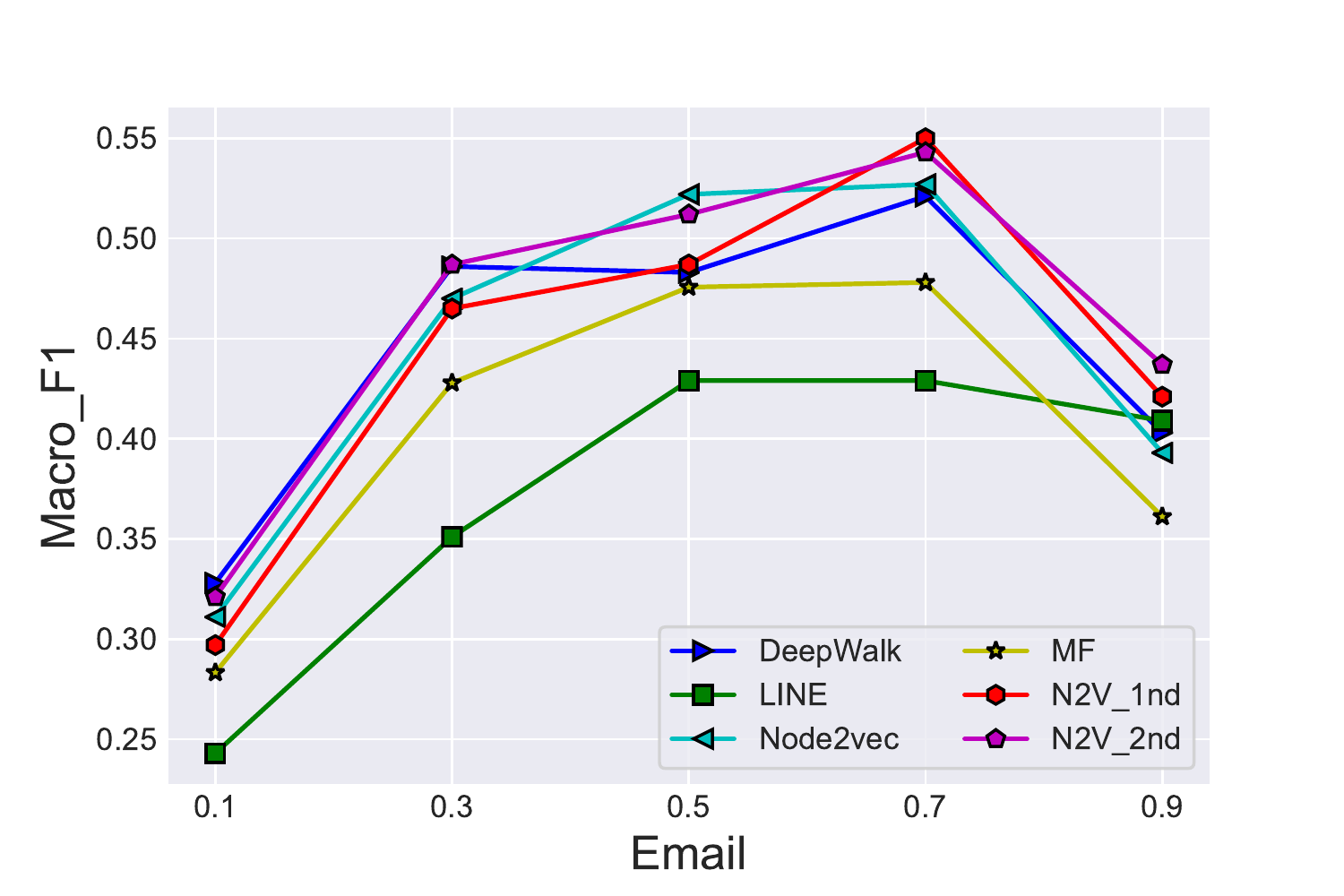}
    }

  \subfigure{
    \includegraphics[height=1.2in, width=1.6in]{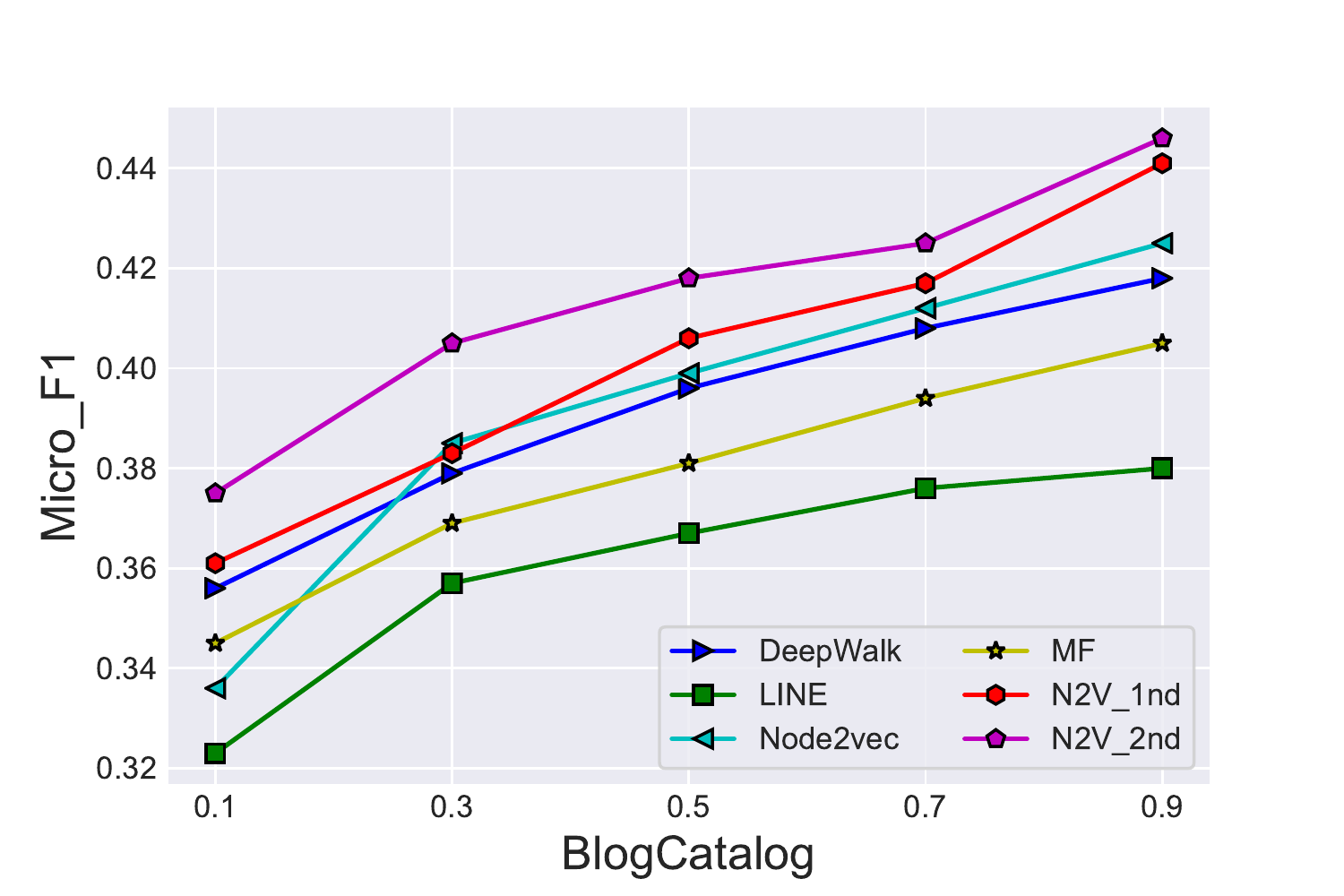}
    \includegraphics[height=1.2in, width=1.6in]{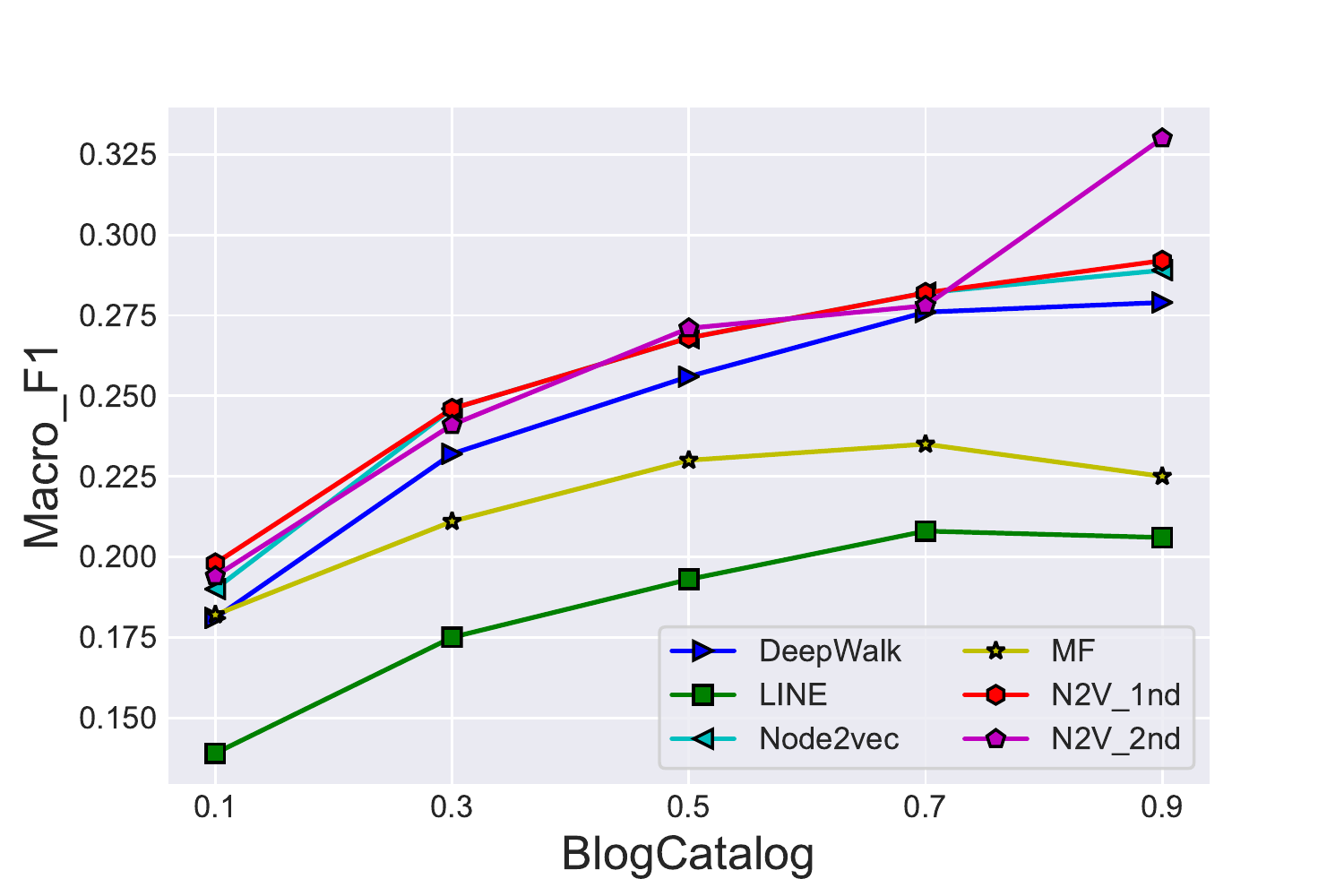}
    }

    \subfigure{
    \includegraphics[height=1.2in, width=1.6in]{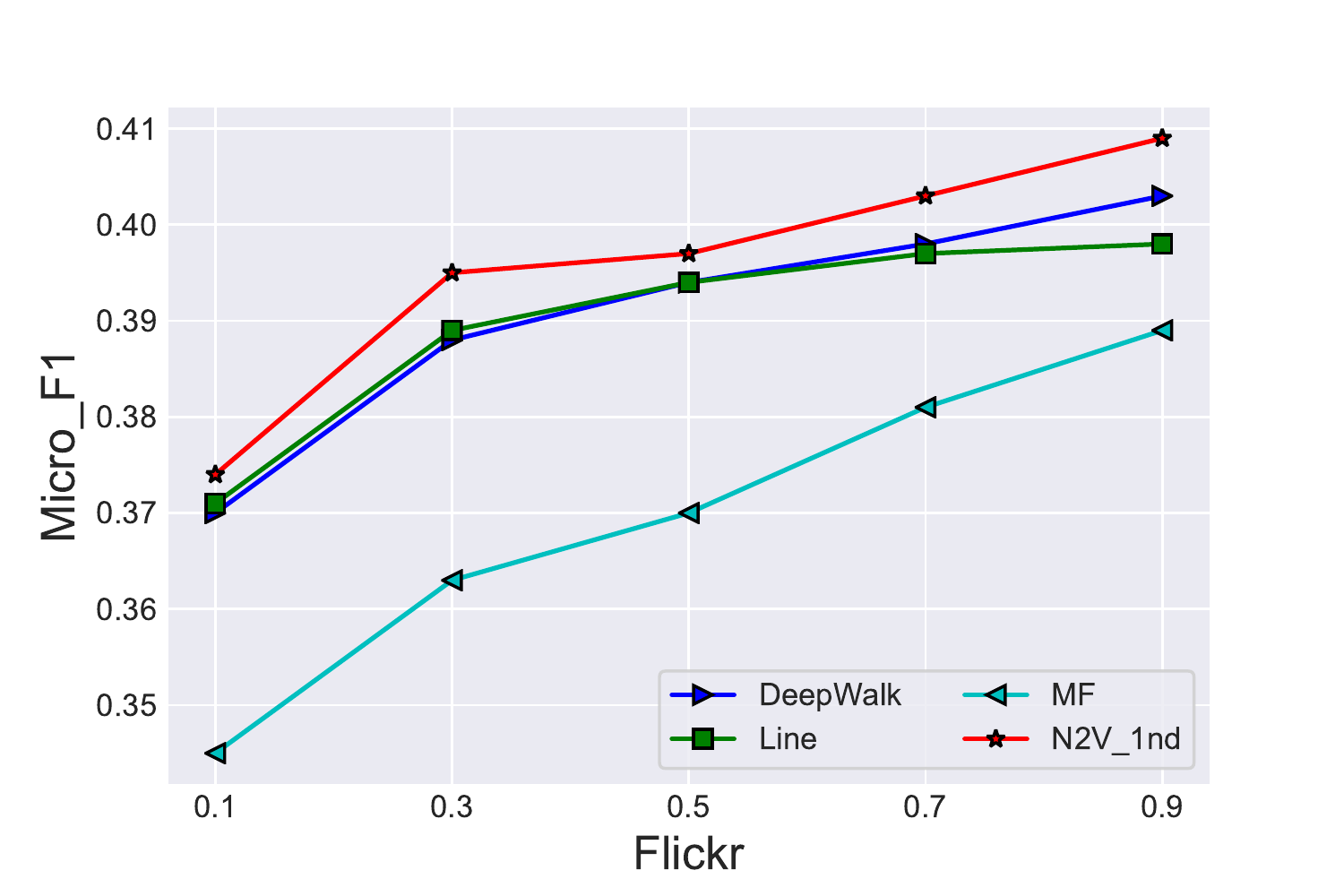}
    \includegraphics[height=1.2in, width=1.6in]{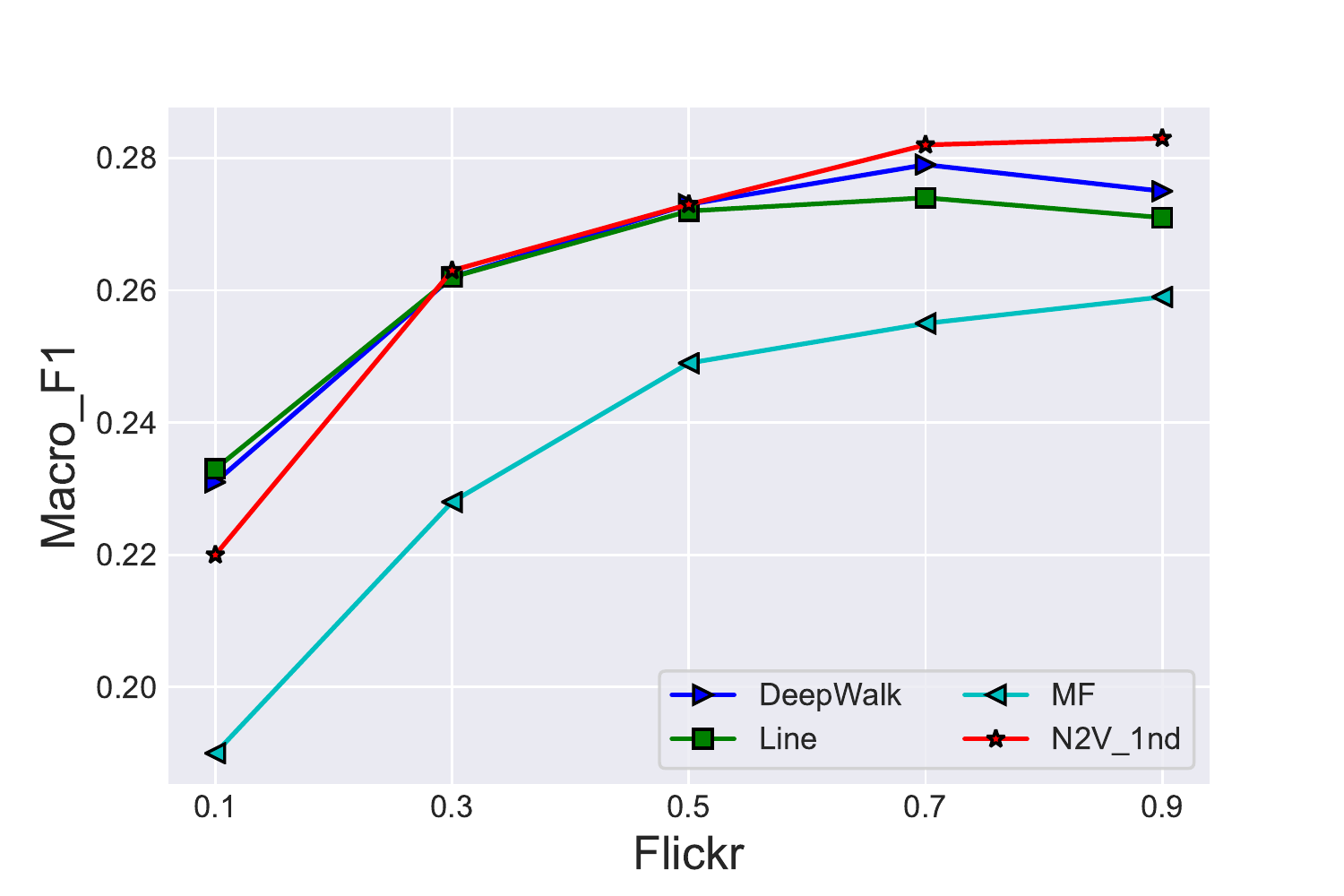}
    }

  \caption{Micro-F1 and Macro-F1 Results in the datasets Email, BlogCatalog, and Flickr.}

\label{fig:classification} 
\end{figure}


\begin{table}
\centering
\caption{Accuracy in the Airport network and the Cora network.}

\begin{tabular}{l | c | c |c|c}\hline
 &Airport-Micro& Airport-Macro & Cora-Micro & Cora-Macro
\\ \hline
DeepWalk &0.475	& 0.451	& 0.903	 &0.707  \\
Node2vec &0.428	&0.44	&0.903	&0.716 \\
LINE	&0.357	& 0.307	&0.9	&0.694 \\
MF	    &0.434	&0.413	&0.906	&0.653 \\
NetMF	&0.482	&0.466	&0.906	&0.705 \\
N2V-1nd	&0.467	& 0.473	& 0.908	& 0.712 \\
N2V-2nd	&\textbf{0.642}	& \textbf{0.655} & \textbf{0.924} & \textbf{0.816}
\\
\hline
\end{tabular}
\label{brazil_airport}
\end{table}

\begin{figure}[!h]
  \centering
    \includegraphics[height=2.8in, width=2.8in]{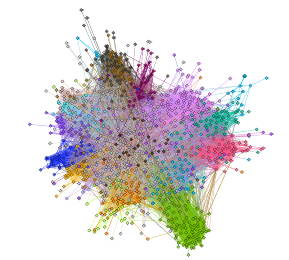}
    \caption{Visualization of the Email network.}
  \label{fig:email_graph} 
\end{figure}

\begin{figure}
  \centering
    \includegraphics[height=2.8in, width=2.8in]{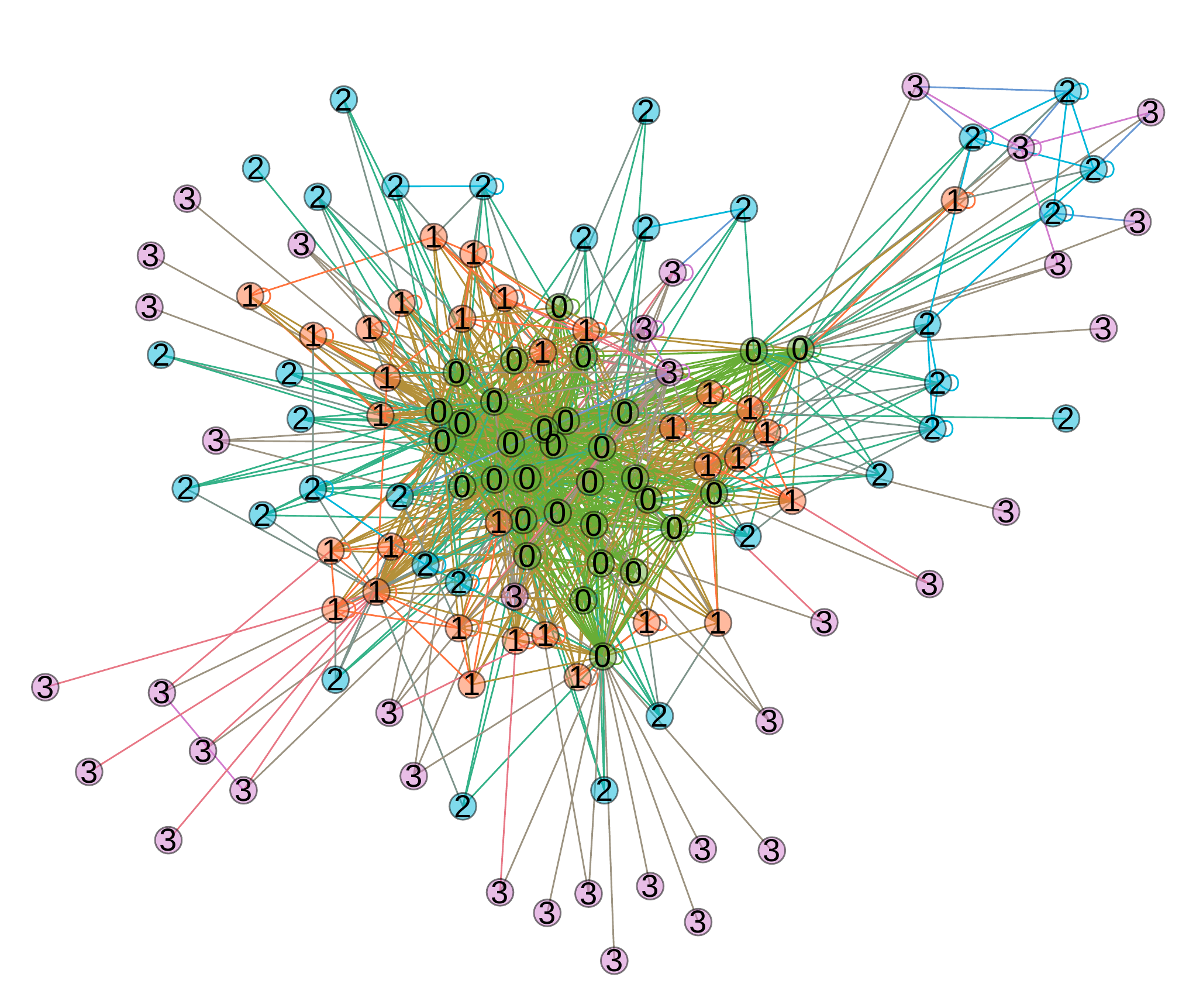}
    \caption{Visualization of the Brazil Airport network.}
  \label{fig:brazil_airport} 
\end{figure}

From the results in the Fig. \ref{fig:classification} and Table \ref{brazil_airport}, it is evident that the proposed method Network2vec out-performs or competitive with baselines in most cases. In the Email network, the Micro-F1 of our model is better than other models. The method LINE performs worst in many cases. Note that the random walks have variance on the performance and the original implementation of Node2vec using a approximating calculation, so Node2vec does not outperform DeepWalk in some cases. In the BlogCatalog dataset \cite{blogflickr2009}, $N2V$ has a Micro-F1 score of 37.4\% while the DeepWalk only has 35.6\% when the training ratio is 10\% . And our model is robust to training ratios and exceeds other methods when more information is obtained. When the training ratio is 90\%, the $N2V$ has a Micro-F1 of 44.6\%, while the Micro-F1 of DeepWalk is 41.8\%.
And $MF$ performs worst in the Flickr network. This may due to the limitation of the pure matrix factorization. The noisy data impacts the performance of $MF$. It is not proper to use Node2vec and $N2V_{2nd}$ due to their high cost of time in larger networks such as the network Flickr. Results on the Flickr of $N2V_{1nd}$ are slightly better than DeepWalk and LINE in Micro-F1, demonstrating the stable performance of $N2V_{1nd}$. From the parameter sensitive experiments, $N2V$ performs better at Macro-F1 when more walks are sampled. However, a better Macro-F1 and Micro-F1 do not always mean a better representation since nodes in some networks are highly connected and cannot be separated absolutely. Sometimes we have to maintain a balance between micro and macro accuracy. In the Brazil Airport network, our method outperforms state-of-the-art methods by a much larger margin. 
We visualize the network and color the nodes by group labels. From the visualization in the Fig. \ref{fig:email_graph} and \ref{fig:brazil_airport}, it can be seen that in the Email network, nodes with the same label tend to be closely connected physically. While in the Airport Network, the group labels reflect the structural similarity of the nodes. The DeepWalk \cite{Deepwalk2014}, LINE \cite{line2015}, Node2vec \cite{node2vec2016} and NetMF \cite{netmf2018wsdm} all fail to produce a meaningful representation, while our method still achieves much better results by around 20\% at most. 
In the Cora network, our method also has a significant improvement compared with previous models. From the perspective of DeepWalk \cite{Deepwalk2014} and Node2vec \cite{node2vec2016}, they try to keep the neighbor proximity while the whole structural semantic information was missing. However, the node representation learned by our model benefits from both global and local information of the whole structure.


\begin{figure}
  \subfigure{
    \includegraphics[height=1.2in, width=1.6in]{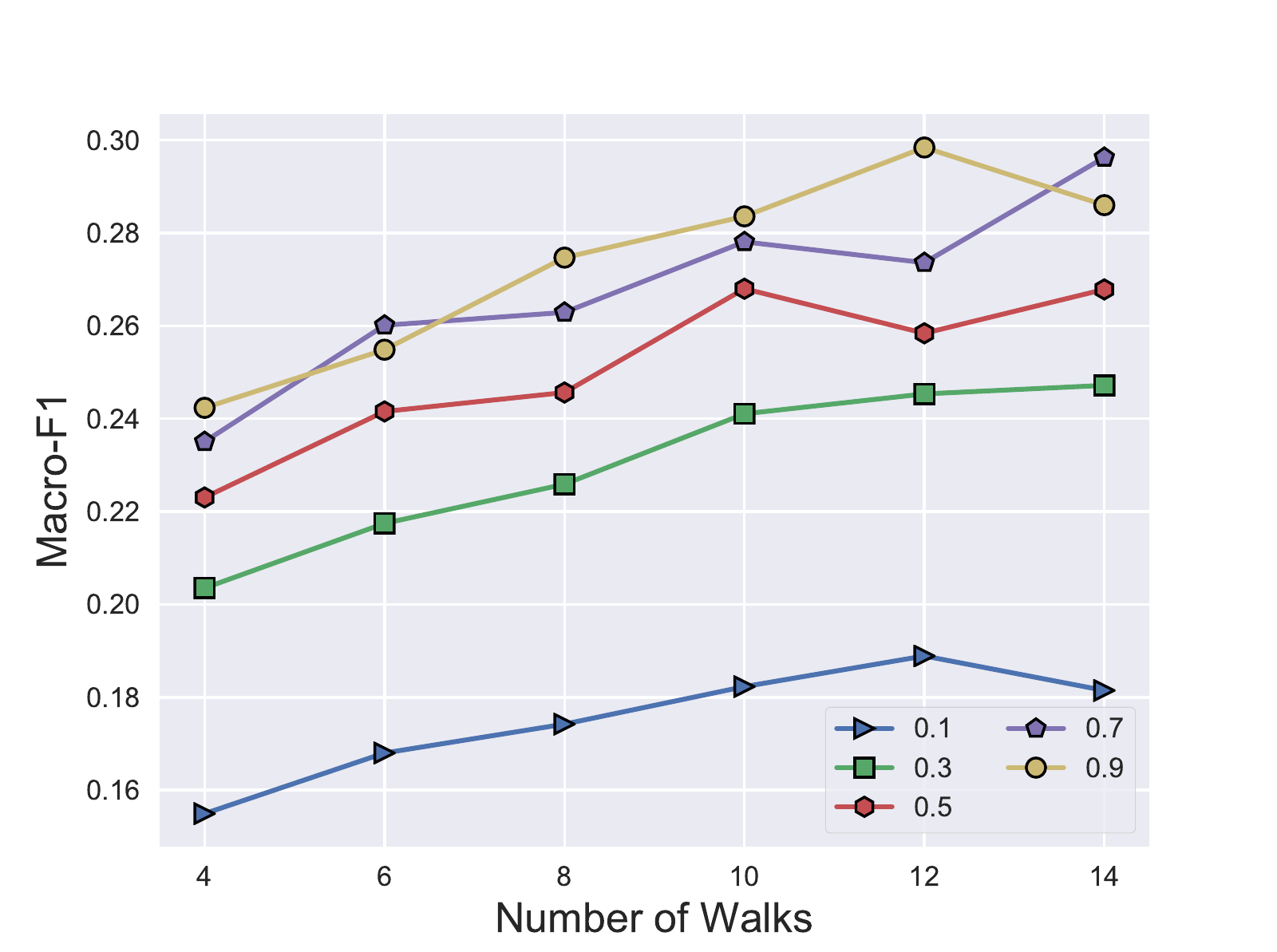}
    }
  \subfigure{
    \includegraphics[height=1.2in, width=1.6in]{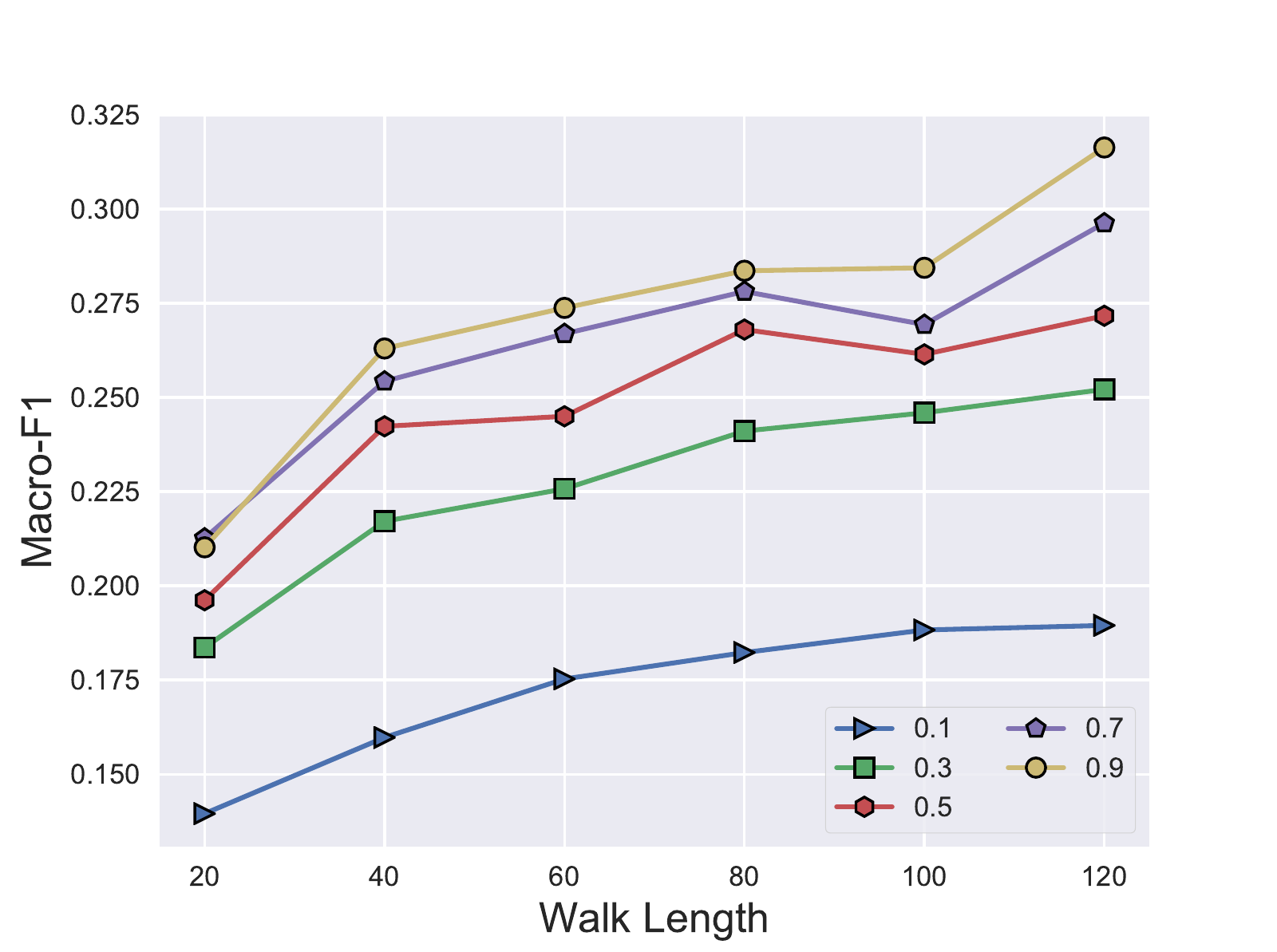}
    }
    \subfigure{
    \includegraphics[height=1.2in, width=1.6in]{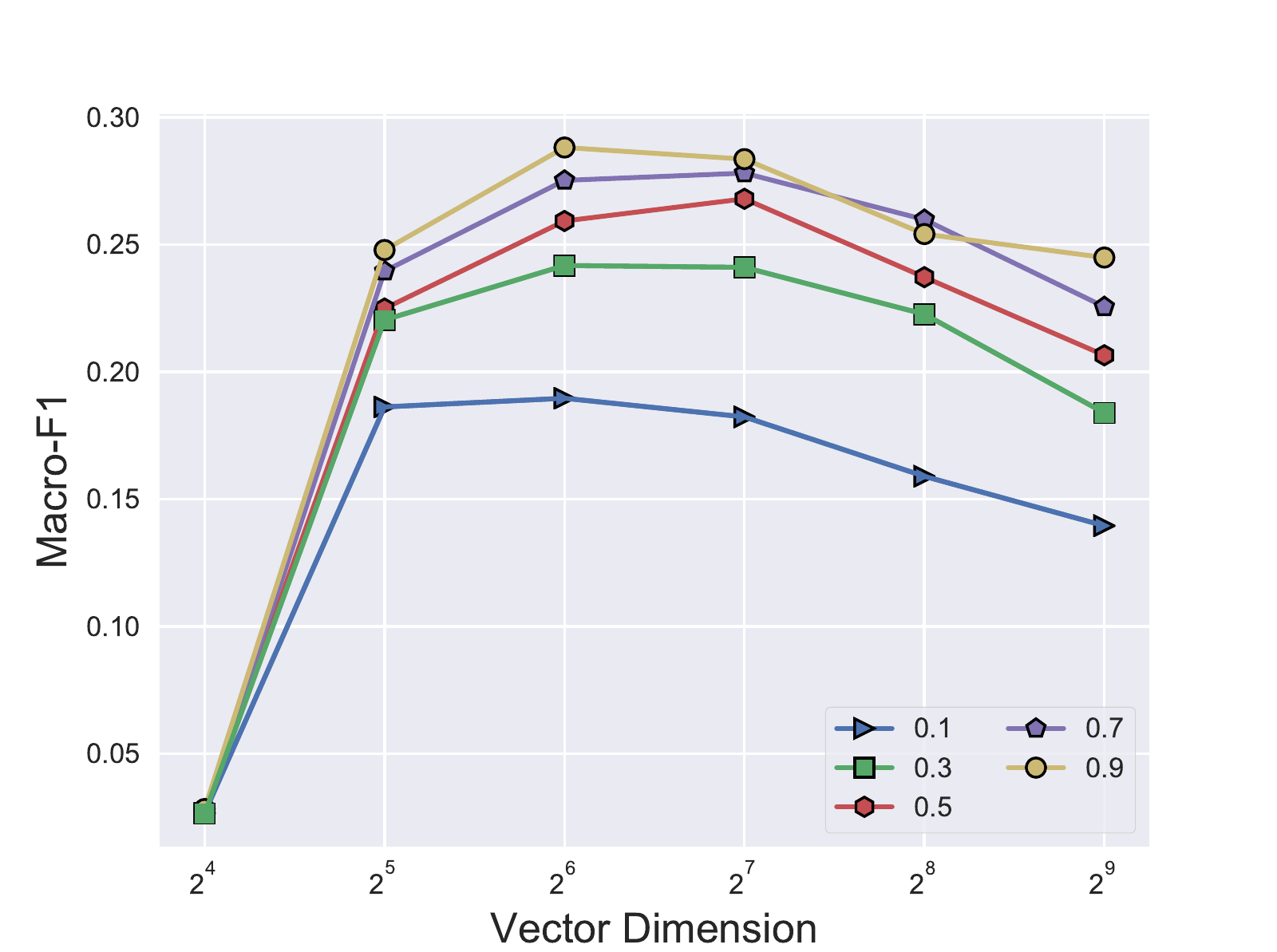}
    }
  \caption{How the parameters of number of walks, walk length, and vector size affect results.}
  \label{fig:parameters} 

\end{figure}

\subsection{\textbf{Link Prediction}}
In the link prediction task, 30\% and 50\% of edges are removed randomly from the network while maintaining the residual network is fully connected as in \cite{node2vec2016}. Then we train embedding models on the left network and make predictions based on training a Logistic Regression classifier. The edge features are represented as the Hadamard product of two nodes' embedding. Since the networks are sparse and the negative samples are much more than positive node pairs. The ratio of positive and negative node pairs is kept to 1:2 to address the unbalance problem. We testify the models on the data sets of the Facebook \cite{node2vec2016} and  Wiki \cite{wikicollective2008data} networks. The Facebook network is a social relationship network collected from participants using a Facebook application in a survey and the data was anonymized. The Wiki network contains 2,405 web pages from 19 categories and 17,981 links between the web pages. As shown in Table \ref{linkprediction}, Our model $N2V$ outperforms DeepWalk\cite{Deepwalk2014} and Node2vec\cite{node2vec2016}. In the Wiki dataset, our model has a significant improvement (by around 10\%) than DeepWalk\cite{Deepwalk2014} and Node2vec\cite{node2vec2016} when 50\% edges are removed. The good performance of our model in the link prediction may due to our model perceives both the local and global information. In link prediction, many real-world networks are not highly separated and links between nodes will have multi-dimensional semantic meaning. Learning from the whole graph view and global information, not just local connection information is conducive to ability of representations produced by the models. The results in link prediction have shown our model has a powerful learning ability in network embedding.

\begin{table}
\centering
\caption{Accuracy in Link Prediction.}
\begin{tabular}{l | c | c |c|c}
\hline
	& Facebook 0.3 &	Facebook 0.5 &	WiKi 0.3 &	WiKi 0.5
\\ \hline
DeepWalk	&0.902	& 0.860	& 0.807	 & 0.777 \\
Node2vec &0.892	&0.830	&0.802	&0.752 \\
LINE	&0.906	&0.880	&0.729	&0.780 \\
MF	&0.895	&0.868	&0.786	&0.778 \\
NetMF	&0.91	&0.871	&0.810	&0.805 \\
N2V-1nd	&0.926	&0.890	& 0.880	&\textbf{0.873} \\
N2V-2nd	&\textbf{0.931}	& \textbf{0.902}	&\textbf{0.882}	&0.866
\\
\hline
\end{tabular}

\label{linkprediction}
\end{table}

\subsection{\textbf{Parameter Sensitivity}}
Since our model includes several parameters. We will discover how the main parameters influence the performance of $N2V_{1nd}$ on the BlogCatalog network. The results are shown in Fig. \ref{fig:parameters}.

There is an obvious growing tendency at Macro-F1 when the number of walks and walk length for a node is increased. The accuracy is not always increased with the vector dimension. A moderate dimension size is recommended between 56 and 128.
\subsection{\textbf{Community Visualization}}
According to \cite{SDNE2016structural}, embedding vectors should reserve more information about original network structure as possible. The network embedding can be treated as an encoding process to some extent \cite{survey2017representation}. We use T-SNE \cite{t-sne2014accelerating} to reduce the embedding vectors into a two-dimensional space and check whether the embedding vectors has reserved the original structure information. Two famous networks are introduced to verify the effectiveness of different models.
\begin{figure}[!h]
  \centering
  \subfigure{
    \includegraphics[height=1.2in, width=1.6in]{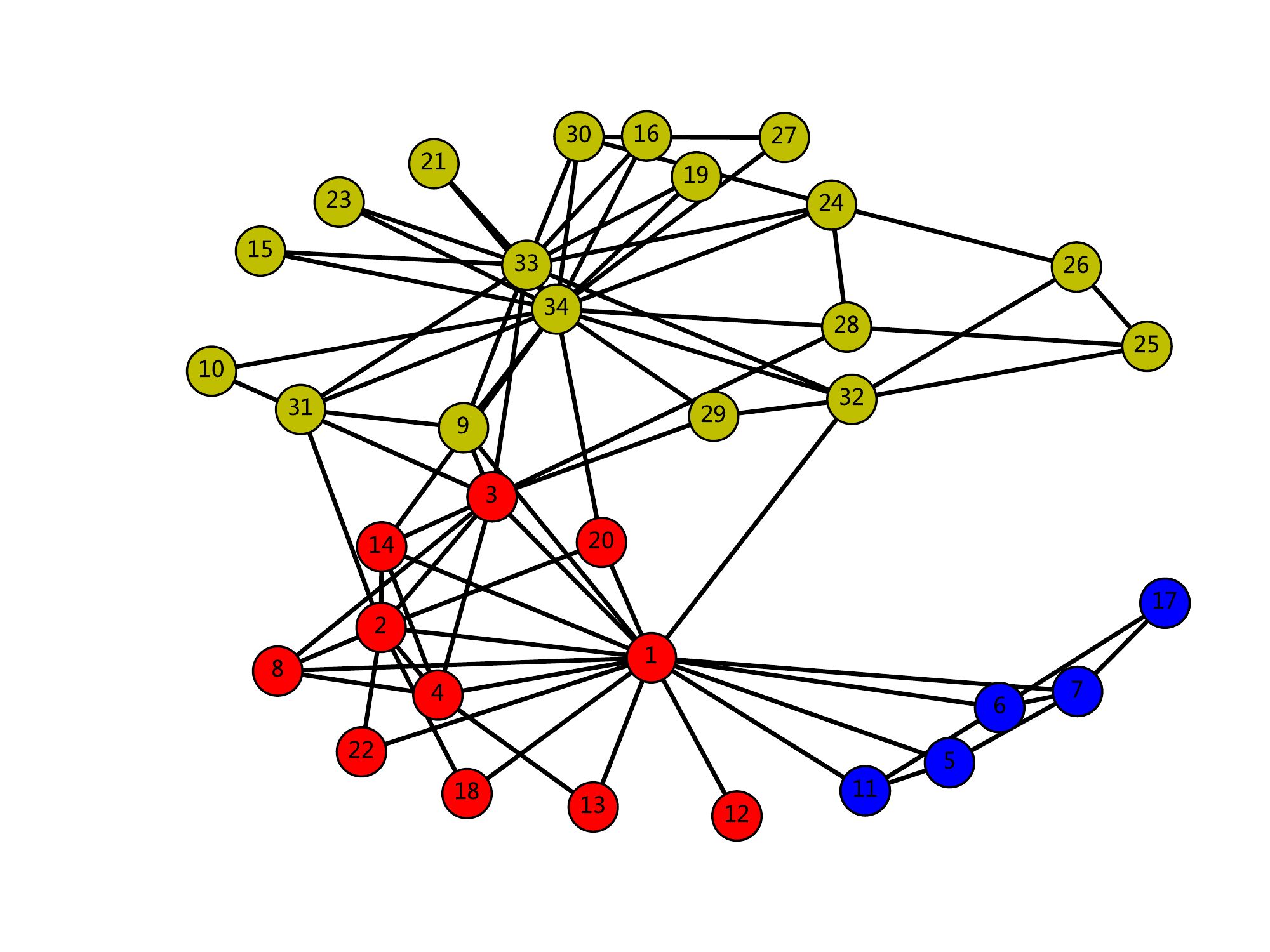}
    \includegraphics[height=1.2in, width=1.6in]{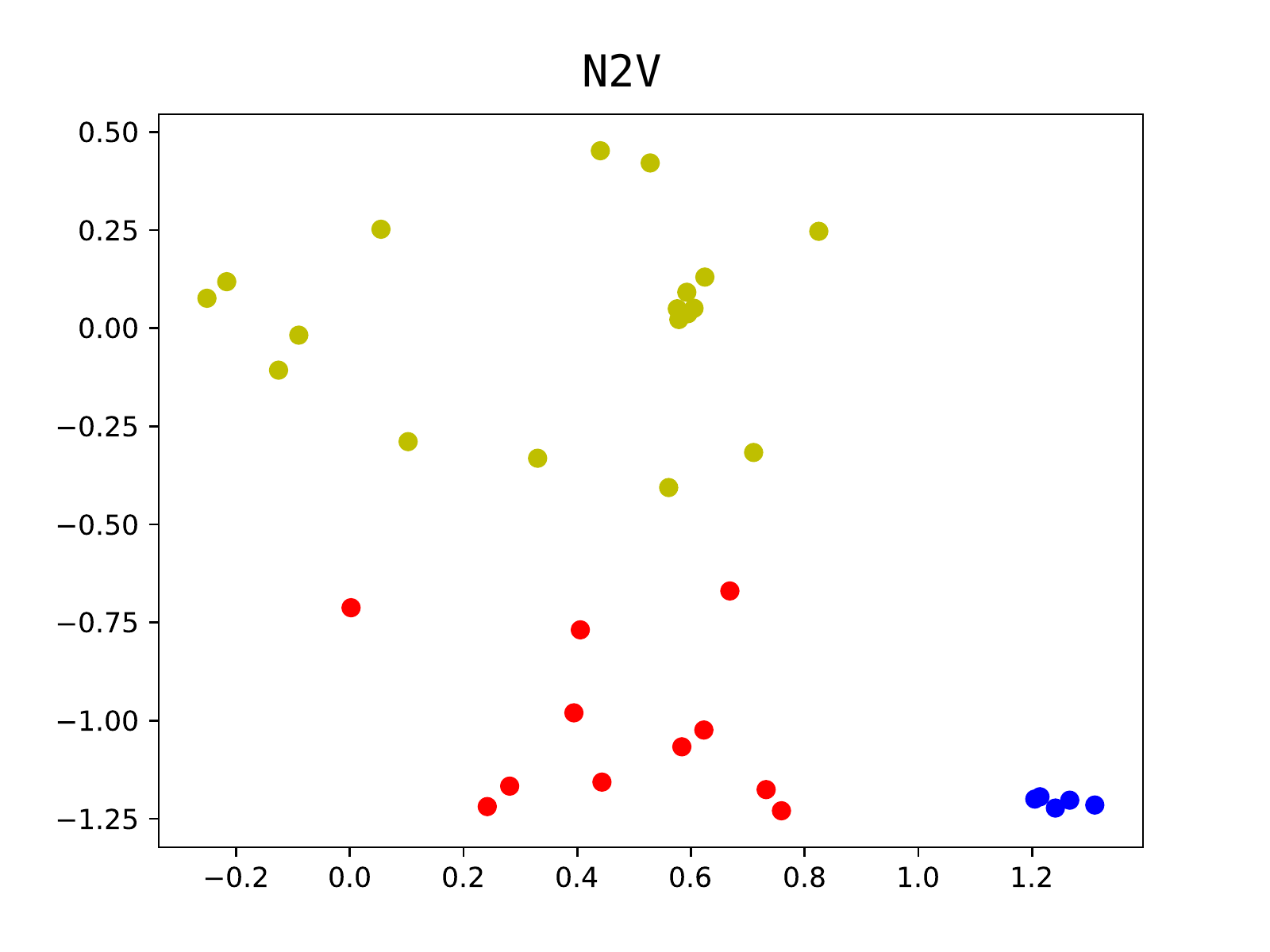}

    }
      \subfigure{
    \includegraphics[height=1.2in, width=1.6in]{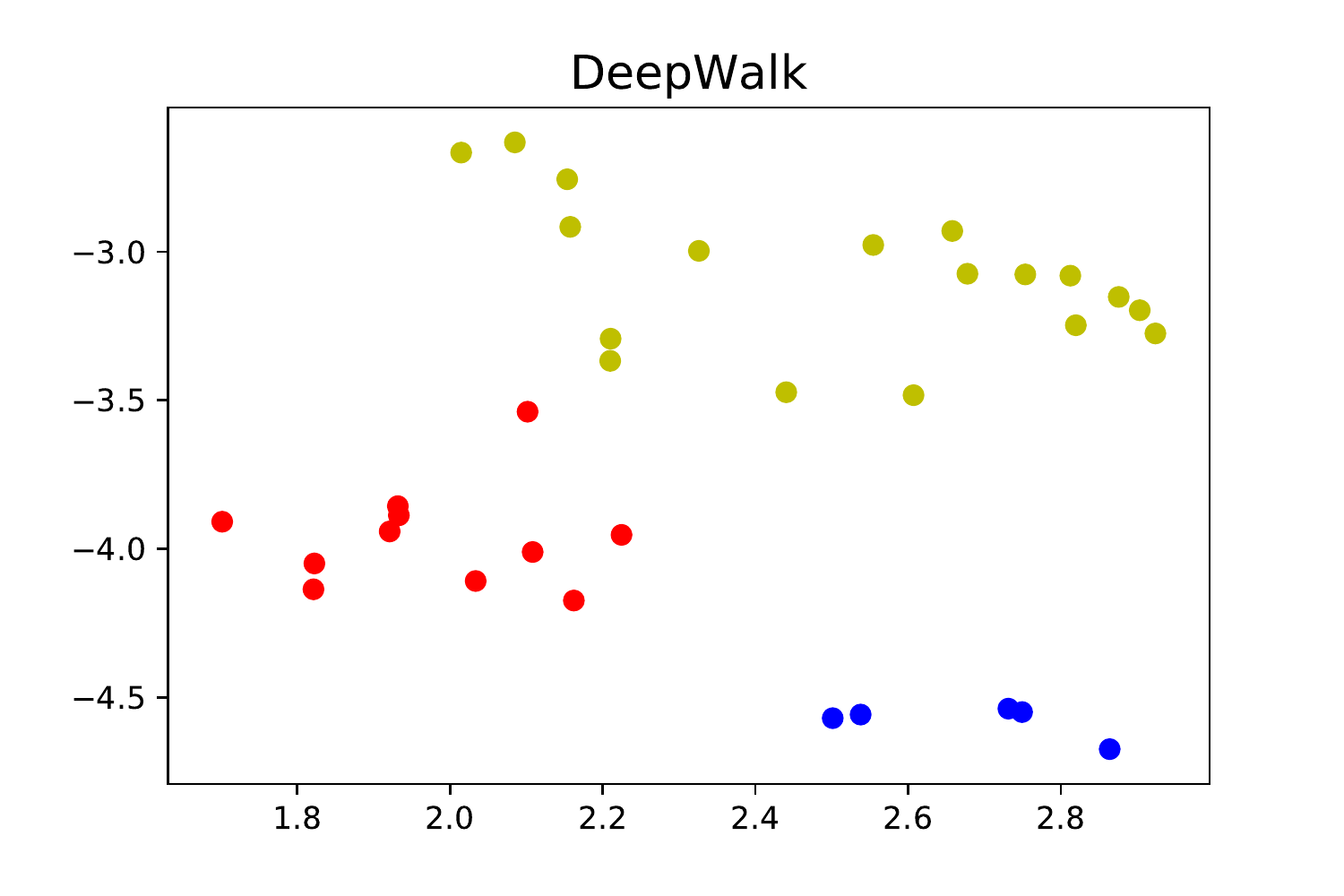}
    \includegraphics[height=1.2in, width=1.6in]{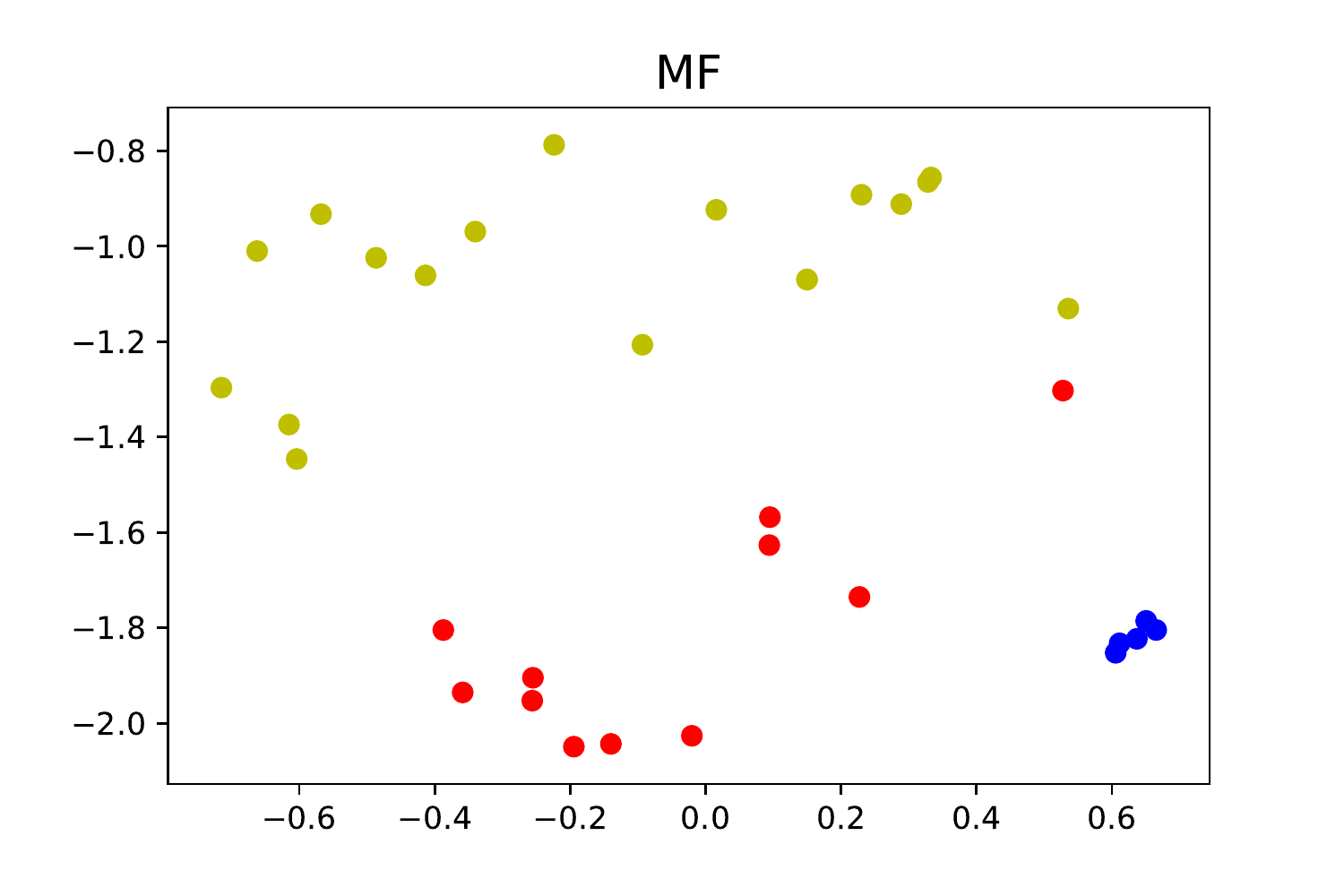}
    }
  \subfigure{
   
    \includegraphics[height=1.2in, width=1.6in]{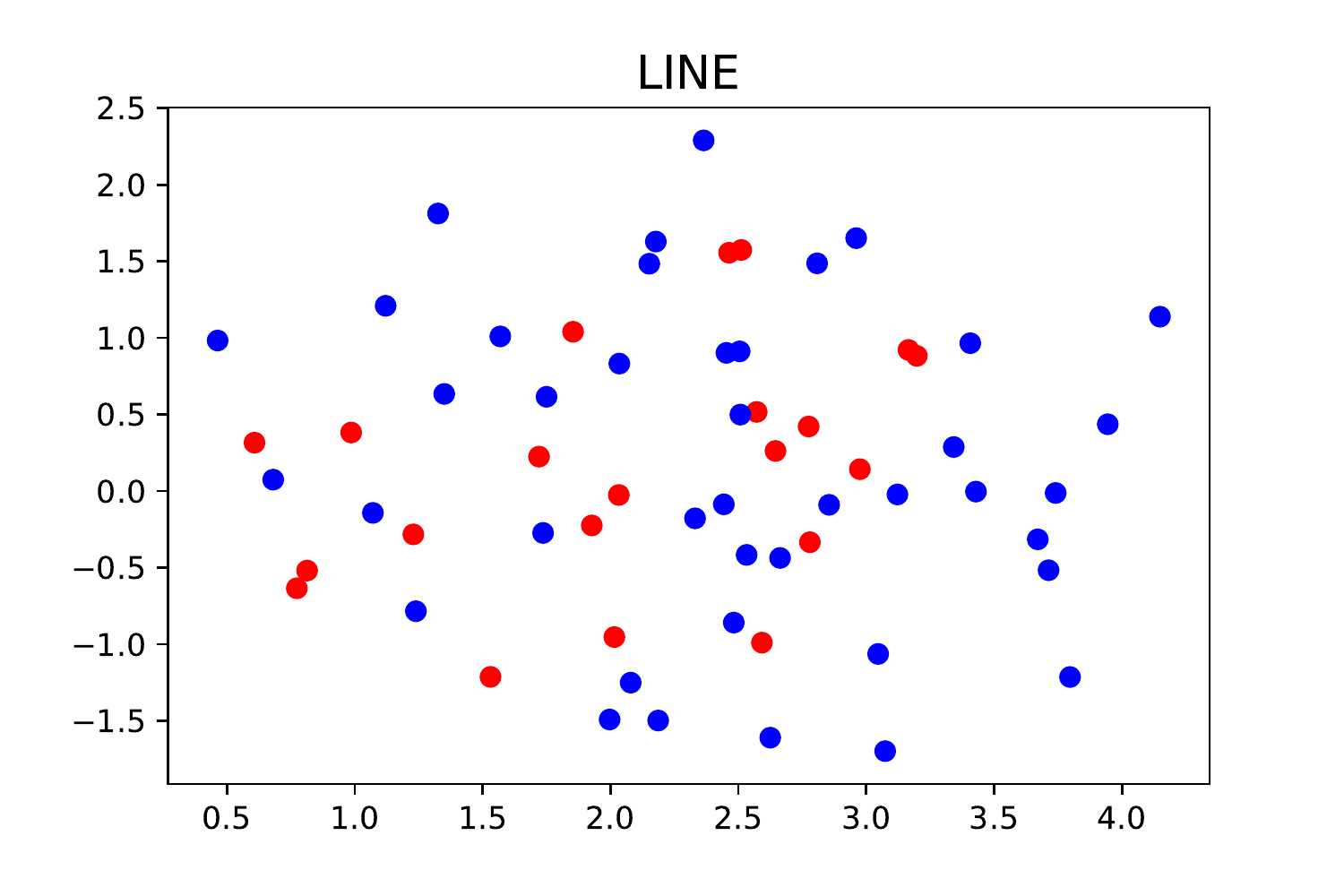}
    \includegraphics[height=1.2in, width=1.6in]{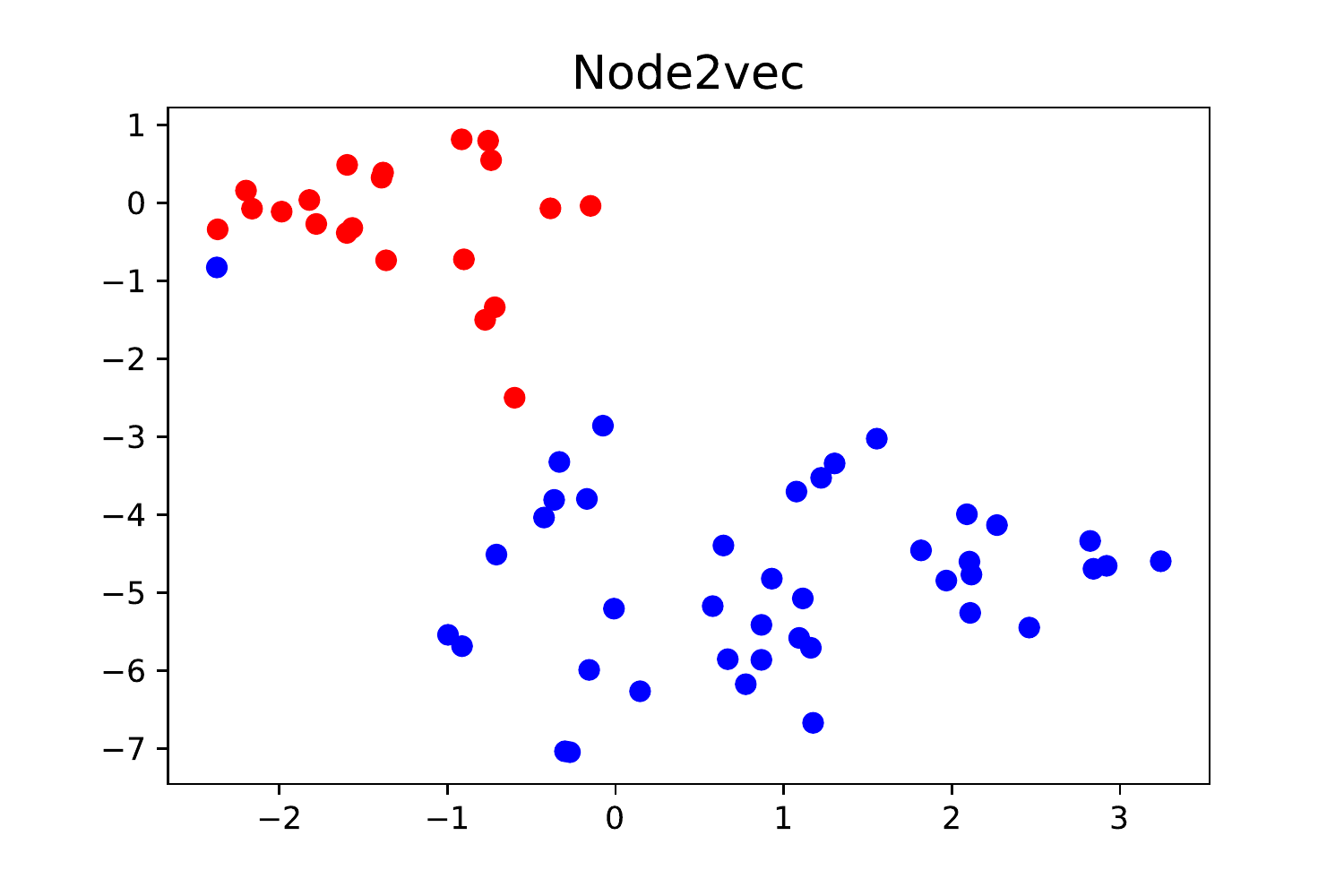}
    }
      \subfigure{
     \includegraphics[height=1.2in, width=1.6in]{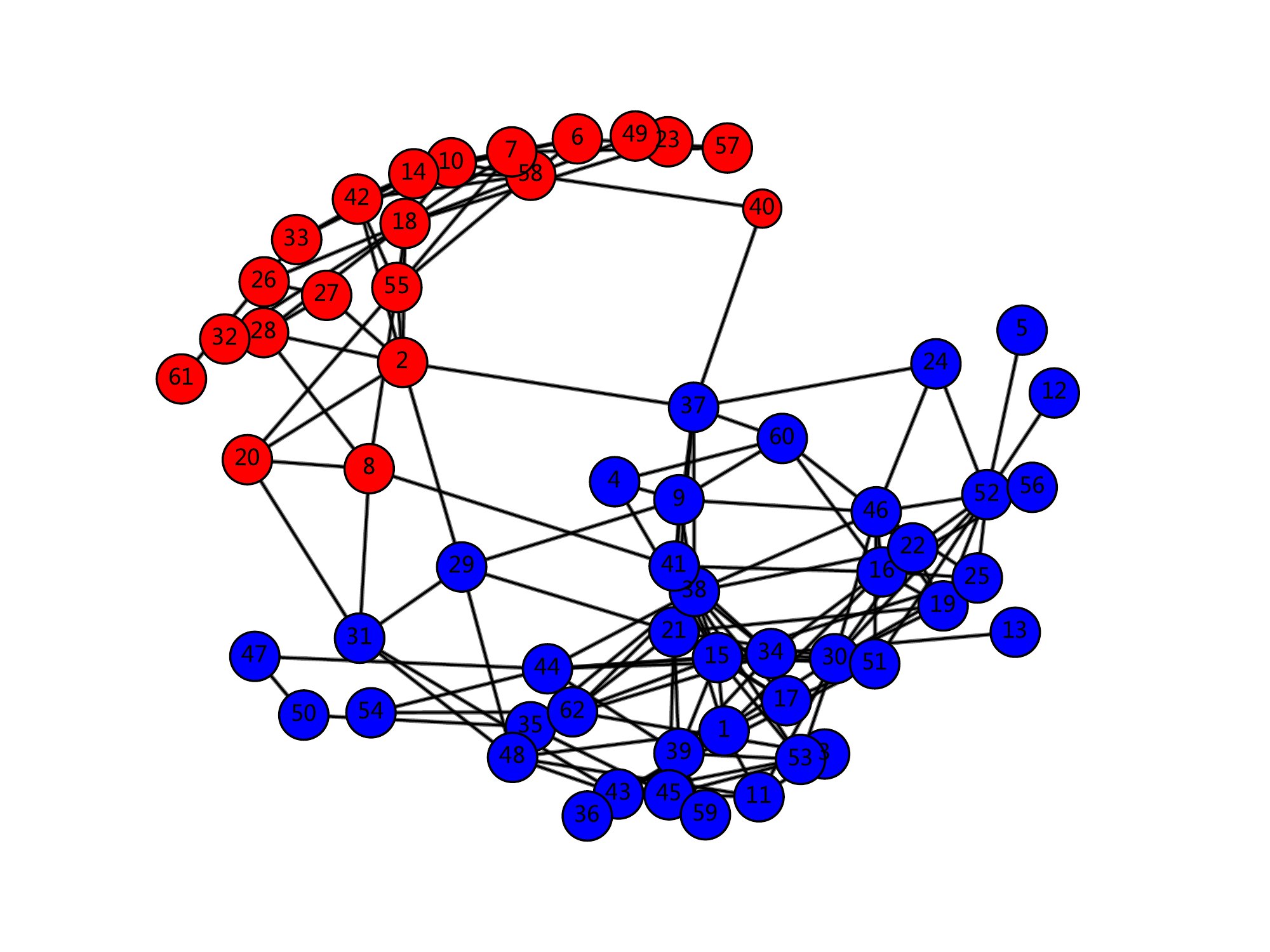}
    \includegraphics[height=1.2in, width=1.6in]{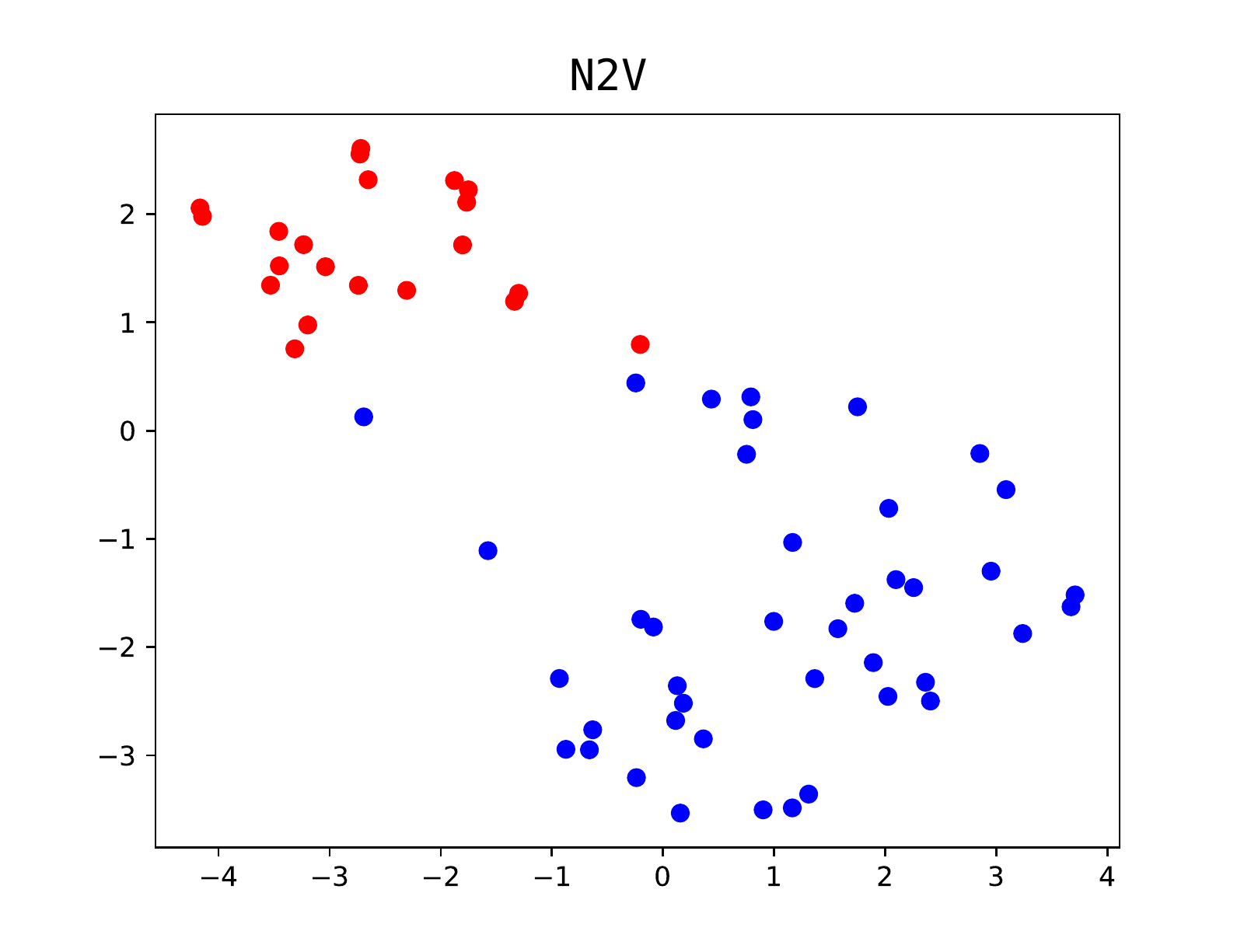}
    }
  \caption{Communities of the karate club network and the bottle nose dolphin network, and visualization of their corresponding embedding vectors.}
  \label{fig:community} 
\end{figure}
\textbf{Karate Club Network}: The Zachary’s Karate Club Network is one of the widely used benchmarks for community detection and validation. Two communities are formed as people surrounded by the coach and the managers separately. \textbf{Bottlenose Dolphins Network}: Some bottlenose dolphins live together in New Zealand at first. These dolphins are generally divided into two groups later due to unknown reasons. Results of different models are shown in Fig. \ref{fig:community}.

It can be seen that the structure of network encoded from embedding space and physical space is surprisingly consistent in our model. In the Dolphin Network, the performance of Node2vec is similar to our model in visualization expect for one blue point. LINE has the worst performance might because the model is more suitable for large graphs. In small networks, nodes are easily connected by the first-order or second-order relationship. The goal of network representation is not just dividing the node into groups, but also keeping physical information of nodes as connectors or contact person between communities. In the Karate Network, the DeepWalk model tends to strongly separate the nodes into groups based on local information. However, our model keeps the local structure as well as perceive the whole network information.

\subsection{\textbf{Training Iterations and Time}}

\begin{table}
\centering
\caption{Training time in different networks.}
\begin{tabular}{l |c | c | c|c } \hline
& Email & Wiki	& BlogCatalog & Flickr\\ \hline
DeepWalk&	16s&	75s	&446s	&2642s\\
LINE&	95s&	93s&	646s&	2611s\\ 
NetMF&	3.6s &	10s	 & 112s	& -\\ \hline
N2V &	5s	& 8s	&82s	&963s \\
\hline\end{tabular}
\label{trainingtime}
\end{table}

$N2V$ can learn a stable model within 10 epochs. Then we compare the training time of $N2V$ with baselines. The parallel threads are set to 10. Training iterations are set to 10. As shown in Table \ref{trainingtime}, compared with DeepWalk, our model is faster than DeepWalk in all the datasets. While the NetFM is also mentionable competitive method. The experiments run on ThinkStation Tower Workstation with Intel Xeon E5-2620 series processors.

\section{Conclusion and Discussion}
From the above experiments, we can conclude that network embedding as pure matrix factorization is not as well as DeepWalk or Node2vec. However, after applying some tricks, the performance can be highly improved. The embedding space and the physical space are fully mapped and have an algebraic relationship. The proposed network embedding model is a flexible and lightweight with a mathematical basic. $Network2vec$ outperforms the previous models on different datasets in the task of node classification. In the link prediction task, our model significantly improves the accuracy. The method $Network2Vec$ enriches the methods of the unsupervised network embedding methods and provides a new perspective to deal with the node representation. Also, $Network2Vec$ is scalable since there are several indicators existed to estimate similarity or distance matrix. The proposed method is potential to be extended in many ways, for example cooperating text and tag information into models. Some other statistical indicators can also be involved in network embedding.

\section*{Acknowledgment}
This  work  was supported by the Science and Technology Program of Guangzhou under Grant No. 201802010025, the Industry-University Research Fund Project of Guangzhou under Grant No. 2019PT103, and the IBM-CSC Young Data Scientists Program.

\bibliography{main.bib}
\bibliographystyle{IEEEtran}

\end{document}